\documentclass[sigconf]{acmart}

\usepackage{balance}
\def\BibTeX{{\rm B\kern-.05em{\sc i\kern-.025em b}\kern-.08emT\kern-.1667em\lower.7ex\hbox{E}\kern-.125emX}}

\setcopyright{none}
\renewcommand\footnotetextcopyrightpermission[1]{}

\usepackage{longtable}
\usepackage{afterpage}
\usepackage{placeins}
\usepackage{rotating}
\usepackage{array}
\usepackage{todonotes}
\usepackage{xspace}
\usepackage[english]{babel}
\usepackage{soul}
\usepackage{flushend}
\usepackage{listings}
\usepackage{amsfonts}
\usepackage{amsmath}
\usepackage{amssymb}
\usepackage{subcaption}
\usepackage{multirow}

\def\nespace{\hskip\fontdimen2\font\relax}

\DeclareMathOperator*{\argmax}{arg\,max}
\usepackage{cleveref}
\Crefformat{figure}{Fig.~#2#1#3}
\Crefmultiformat{figure}{Figs.~#2#1#3}{ and~#2#1#3}{, #2#1#3}{ and~#2#1#3}

\makeatletter
\patchcmd{\@makecaption}
  {\scshape}
  {}
  {}
  {}

\newcommand*{\da@rightarrow}{\mathchar"0\hexnumber@\symAMSa 4B }
\newcommand*{\da@leftarrow}{\mathchar"0\hexnumber@\symAMSa 4C }
\newcommand*{\xdashrightarrow}[2][]{%
  \mathrel{%
    \mathpalette{\da@xarrow{#1}{#2}{}\da@rightarrow{\,}{}}{}%
  }%
}
\newcommand{\xdashleftarrow}[2][]{%
  \mathrel{%
    \mathpalette{\da@xarrow{#1}{#2}\da@leftarrow{}{}{\,}}{}%
  }%
}
\newcommand*{\da@xarrow}[7]{%
  \sbox0{$\ifx#7\scriptstyle\scriptscriptstyle\else\scriptstyle\fi#5#1#6\m@th$}%
  \sbox2{$\ifx#7\scriptstyle\scriptscriptstyle\else\scriptstyle\fi#5#2#6\m@th$}%
  \sbox4{$#7\dabar@\m@th$}%
  \dimen@=\wd0 %
  \ifdim\wd2 >\dimen@
    \dimen@=\wd2 %
  \fi
  \count@=2 %
  \def\da@bars{\dabar@\dabar@}%
  \@whiledim\count@\wd4<\dimen@\do{%
    \advance\count@\@ne
    \expandafter\def\expandafter\da@bars\expandafter{%
      \da@bars
      \dabar@ 
    }%
  }%
  \mathrel{#3}%
  \mathrel{%
    \mathop{\da@bars}\limits
    \ifx\\#1\\%
    \else
      _{\copy0}%
    \fi
    \ifx\\#2\\%
    \else
      ^{\copy2}%
    \fi
  }%
  \mathrel{#4}%
}
\makeatother

\newcommand{\projname}{\textsc{Poirot}\xspace}

\newcolumntype{M}[1]{>{\centering\arraybackslash}p{#1}}

\usepackage{enumitem}
\setlist[itemize,1]{leftmargin=1.5\parindent, itemsep=0ex, topsep=0.5ex, %
}
\setlist[enumerate,1]{leftmargin=2\parindent, itemsep=0ex, topsep=0.5ex, %
}

\begin{document}

\fancyhead{}

\title[\projname: Aligning Attack Behavior with Kernel Audit Records]{\projname: Aligning Attack Behavior with Kernel Audit Records for Cyber Threat Hunting}

 \author{Sadegh M. Milajerdi}
 \email{smomen2@uic.edu}
 \affiliation{
 \institution{University of Illinois at Chicago}
 }
 \author{Birhanu Eshete}
 \email{birhanu@umich.edu}
 \affiliation{
 \institution{University of Michigan-Dearborn}
 }
 \author{Rigel Gjomemo}
 \email{rgjome1@uic.edu}
 \affiliation{
 \institution{University of Illinois at Chicago}
 }
 \author{V.N. Venkatakrishnan}
 \email{venkat@uic.edu}
 \affiliation{
 \institution{University of Illinois at Chicago}
 }

 \renewcommand{\shortauthors}{Sadegh M. Milajerdi et al.}

\begin{abstract}
Cyber threat intelligence (CTI) is being used to search for indicators of attacks that might have compromised an enterprise network for a long time without being discovered.
To have a more effective analysis, CTI open standards have incorporated descriptive relationships showing how the indicators or observables are related to each other.
However, these relationships are either completely overlooked in information gathering or not used for threat hunting.
In this paper, we propose a system, called \projname, which uses these correlations to uncover the steps of a successful attack campaign.
We use kernel audits as a reliable source that covers all causal relations and information flows among system entities and model threat hunting as an inexact graph
pattern matching problem. Our technical approach is  based on a novel similarity metric which assesses an alignment between a query graph constructed out of CTI correlations and a  provenance graph constructed out of kernel audit log records.
We evaluate \projname on publicly released real-world incident reports as well as reports of an  adversarial engagement designed by
DARPA, including ten distinct attack campaigns against  different OS platforms such as Linux, FreeBSD, and Windows. 
Our evaluation results show that \projname is capable of searching inside graphs containing millions of nodes and pinpoint the attacks in a few minutes, and the results serve to illustrate that  CTI correlations could be used as robust and reliable artifacts for threat hunting.

\end{abstract}

\keywords{Cyber Threat Hunting, Cyber Threat Intelligence, Indicator of Compromise, Graph Alignment, Graph Pattern Matching}
\maketitle

\textbf{Preprint.} The final version of this paper is going to appear in the 
ACM SIGSAC Conference on Computer and Communications Security (CCS'19), November 11-15, 2019, London, United Kingdom. 

\section{Introduction}\label{sec:intro}

When Indicators of Compromise (IOCs) related to an advanced persistent threat (APT) detected inside an organization are released, a common question that emerges among enterprise security analysts is if their enterprise has been the target of that APT. This process is  commonly known as \textit{Threat Hunting}. Answering this question with a high level of confidence often requires  lengthy and complicated searches and analysis over host and network logs of the enterprise, recognizing entities that appear in the IOC descriptions among those logs and finally assessing the likelihood that the specific APT successfully infiltrated the enterprise. %

In general, threat hunting inside an enterprise presents several challenges:
\begin{itemize}

\item {\em Search at scale}: 
To remain under the radar, an attacker often performs the attack steps over long periods (weeks, or in some cases, months).  Hence, it is necessary to design an approach that can link related IOCs together even if they are conducted over a long period of time. To this end, the system should be capable of searching among 
millions of log events (99.9\% of which often correspond to benign activities).

\item {\em Robust identification and linking of threat-relevant entities}: 
Threat hunting
must be sound in identifying whether 
an attack campaign has affected a system, even though the attacker might have mutated the artifacts like file hashes and IP addresses to evade detection.
Therefore,
a robust approach should not merely look for
matching IOCs in isolation, but uncover the entire threat scenario, which is harder for an attacker to mutate. 

\item {\em Efficient Matching}:
For a cyber analyst to understand and react to a threat incident in a timely fashion, the approach must efficiently conduct the search and not produce many false positives so that appropriate cyber-response operations can be initiated in a timely fashion.

\end{itemize}

 Commonly, knowledge about the malware employed in APT campaigns is published in cyber threat intelligence (CTI) reports and is presented in a variety of forms such as natural language,  structured, and semi-structured form. To facilitate the smooth exchange of CTI in the form of IOCs and enable characterization of adversarial techniques, tactics, and procedures (TTPs), the security community has adopted open standards such as OpenIOC \cite{open-ioc}, STIX \cite{stix}, and MISP \cite{misp}.
To provide a better overview of attacks, these standards often incorporate descriptive relationships showing how indicators or observables are related to each other \cite{misp-correlations}.

However, a vast majority of the current threat hunting approaches operates only over fragmented views of cyber threats\cite{loki,redline}, such as signatures (e.g., hashes of artifacts), suspicious file/process names, and IP addresses (domain names), or by using heuristics such as timestamps to correlate suspicious events \cite{pei2016hercule}. These approaches are useful but have limitations, such as (i) lacking the precision to reveal the complete picture as to how the threat unfolded especially over long periods (weeks, or in some cases, months), 
(ii) being susceptible to false signals when adversaries use legitimate-looking names (like svchost in Windows) to make their attacks indistinguishable from benign system activities, and (iii)
relying on low-level signatures, which makes them ineffective when attackers  update or re-purpose ~\cite{nsa-tools-china,nsa-tools-china1} their tools or change their signatures (IP addresses or hash values) to evade detection.
To overcome these limitations and build a robust detection system, the correlation among IOCs must be taken into account. In fact, 
the relationships between IOC artifacts contain essential clues on the {\em behavior} of the attacks inside a compromised system, which is
tied to attacker goals and  is, therefore,  more difficult to change \cite{kolbitsch2009effective,zong2015behavior}.

This paper formalizes the threat hunting problem from CTI reports and IOC descriptions, develops a rigorous approach for deriving the confidence score that indicates the likelihood of success of an attack campaign, and describes a system called \projname that implements this approach. In a nutshell, given a graph-based representation of IOCs and relationships among them that expresses the overall behavior of an APT, which we call  a {\em query graph}, our approach efficiently finds an embedding of this {\em query graph} in a much larger {\em provenance graph}, which contains a representation of kernel audit logs over a long period of time. 
Kernel audit logs 
are free of unauthorized tampering as long as system's kernel is not compromised, and reliably contain relationships between system entities (e.g., processes, files, sockets, etc.), 
in contrast to   its alternatives (e.g., firewall, network monitoring, and file access
logs) which provide partial information. We assume that to maintain the integrity of kernel audit logs,  a real-time kernel audit storage on a separate and secure log server is used as a precaution against log tampering. 

More precisely, we formulate threat hunting as a graph pattern matching (GPM) problem  searching for causal dependencies or information flows among system entities  that are similar to those described in the {\em query graph}.
To be robust against evasive attacks (e.g., mimicry attacks \cite{mimicry,asiaccs08}) which aim to influence the matching, we prioritize flows based on the cost they have for an attacker to produce.
Given the NP-completeness of the graph matching problem \cite{de2009subgraph}, we propose an approximation function and a novel similarity metric to assess an alignment between the \textit{query} and \textit{provenance} graph.

We
test \projname's effectiveness and efficiency using three different datasets, particularly, red-team/blue-team adversarial engagements performed by DARPA Transparent Computing (TC) program \cite{tc_github}, publicly available real-world incident reports, and attack-free activities generated by ordinary users. In addition, we simulate several attacks from real-world scenarios in a controlled environment and compare \projname with other tools that are currently used to do threat hunting. We show that \projname outperforms these tools. 
We have implemented different kernel log parsers for Linux, FreeBSD, and Windows, and our evaluation results show that \projname can search inside graphs containing millions of nodes and pinpoint the attacks in a few minutes.

This paper is organized as follows: 
Related work appears in ~\cref{sec:related}.
We present an overall architecture of \projname in ~\cref{sec:approach_overview}.
In ~\cref{sec:approach}, we provide the formal details of the graph alignment algorithm.
~\Cref{sec:eval} discusses the evaluation, and we conclude in  ~\cref{sec:conclusion}. 
\vspace{-1em}

\section{Related Work}\label{sec:related}

{\bf Log-based Attack Analytics.}
Opera et al. \cite{oprea2015detection} leverage DNS or web proxy logs for detecting early-stage infection in an enterprise.
{\sc Disclosure} \cite{bilge2012disclosure} extracts statistical features from NetFlow logs to detect botnet C\&C channels.
DNS logs have also been extensively used \cite{antonakakis2011detecting,antonakakis2012throw} for detecting malicious domains.
{\sc Hercule} \cite{pei2016hercule} uses community detection to reconstruct attack stages by correlating logs coming from multiple sources.
Similar to \projname, a large body of work uses system audit logs to perform forensic analysis and attack reconstruction\cite{taser2005,forensix,pohly2012hi,liu2018towards}.

{\bf Provenance Graph Explorations.}
The idea to construct a provenance graph from kernel audit logs was introduced by King et al. \cite{king2003backtracking,king2005enriching}. 
The large size and coarse granularity of these graphs have limited their practical use.
However, recent advancements have paved the way for more efficient and effective use of provenance graphs.
Several approaches have introduced compression, summarization, and log reduction techniques \cite{lee2013loggc,xu2016high,depPresRed18} to differentiate worthy events from uninformative ones and consequently reduce the storage size. %
Dividing processes into smaller units is one of the approaches to add more granularity into the provenance graphs, and to this end, researchers have utilized different methods, such as dynamic binary analysis \cite{lee2013high,ma2016protracer}, source code annotation \cite{ma2017mpi}, or modeling-based inference \cite{Ma2015Accurate,kwon18mci,sadegh2018propatrol}.
Additionally, record-and-replay \cite{ji2017rain,ji2018enabling} and parallel execution methods \cite{kwon2016ldx} are  proposed for more precise tracking.
Recent studies have leveraged provenance graphs for different objectives, such as alert triage \cite{hassan2019nodoze}, zero-day attack path identification \cite{sun2018using}, attack detection and  reconstruction \cite{sadegh2019holmes,hossain2017sleuth}. However, the scope of \projname is different from these recent works, since it is focused on {\em threat hunting} and not real-time detection or forensic analysis.  

{\bf Query Processing Systems.}
Prior works have incorporated  novel optimization techniques, graph indexing, and query processing methods  \cite{wang2012efficient,giugno2002graphgrep,sun2012efficient} 
to support timely attack investigations.
SAQL \cite{gao2018saql} is an anomaly query engine that
queries  specified anomalies to identify abnormal behaviors among system  events.
AIQL \cite{gao2018aiql} can be used as a forensic query system that has a domain-specific language for investigating attack campaigns from historical audit logs.
Pasquier et al. \cite{pasquierruntime} propose a query framework, called {\sc CamQuery}, that supports real-time analysis on provenance graphs, to address problems such as data loss prevention, intrusion detection, and regulatory compliance.
Shu et al.  \cite{Shu:2018:TIC:3243734.3243829} also propose a human-assisted query system equipping threat hunters with a suite of potent
new tools. 
These works are orthogonal to \projname and can be used as a foundation to implement our search algorithm.

{\bf Behavior Discovery.}
Extracting malicious behaviors such as information flows and causal dependencies and searching for them as robust indicators have been investigated in prior works.
Christodorescu et al. \cite{christodorescu2007mining} have proposed an approach for mining malware behavior from dynamic traces of that malware's samples.
Similarly, Kolbitsch et al. \cite{kolbitsch2009effective} automatically generate behavior models of malware using symbolic execution.
They represent this behavior as a graph and search for it among the runtime behavior of unknown programs.
On the contrary, \projname does not rely on symbolic expressions but looks for correlations and information flows on the whole system.
TGMiner
\cite{zong2015behavior} is a method to mine discriminative graph patterns from training audit logs
and search for their existence in test data.
The focus of this work
is query formulation instead of pattern query processing, and the authors have used a subsequence matching solution 
\cite{zong2014cloud} for their search, which is different from our graph pattern matching approach.

{\bf Graph Pattern Matching.}
Graph Pattern Matching (GPM) has proved useful in a variety of applications \cite{gallagher2006matching}.
GPM can be defined as a search problem inside a large graph for a subgraph containing similar connections  conjunctively specified in a small query graph.
This problem is NP-complete in the general case \cite{de2009subgraph}.
Fan et. al. \cite{fan2010graph} proposed a polynomial time approach assuming that each connection in the pattern could only be mapped to a path with a
predefined number of hops.
Other works \cite{cheng2008fast,zou2009distance} have tackled the problem by using a sequence of join functions in the vector space.
{\sc NeMa} \cite{khan2013nema} is a neighborhood-based subgraph matching technique based on the proximity of nodes.
In contrast, {\sc G-Ray} and later {\sc Mage}\cite{gray,pienta2014mage} take into account the shape of the query graph and edge attributes and are more similar to our approach, where similar information flows and causal dependencies play a crucial role.
However, these approaches work based on random-walk, which is not reliable against  attackers (with knowledge of the threat-hunting method) who generate fake events (as explained in \cref{sec:score_metric}).
While our graph alignment notions are similar to 
these works, the graph characteristics \projname analyzes present  new challenges  such as being labeled, directed, typed, in the order of millions of nodes, and constructed in an adversarial setting.  
Moreover, many of these related works are looking for a subgraph that contains exactly one alignment for each node and each edge of the query graph and cannot operate in a setting where there might not be an alignment for certain nodes or edges.
As a result, we develop a new best-effort matching technique  aimed at tackling these challenges.

\section{Approach Overview}\label{sec:approach_overview}

\begin{figure}[t]
 \begin{center}
 \includegraphics[width=.95\columnwidth]{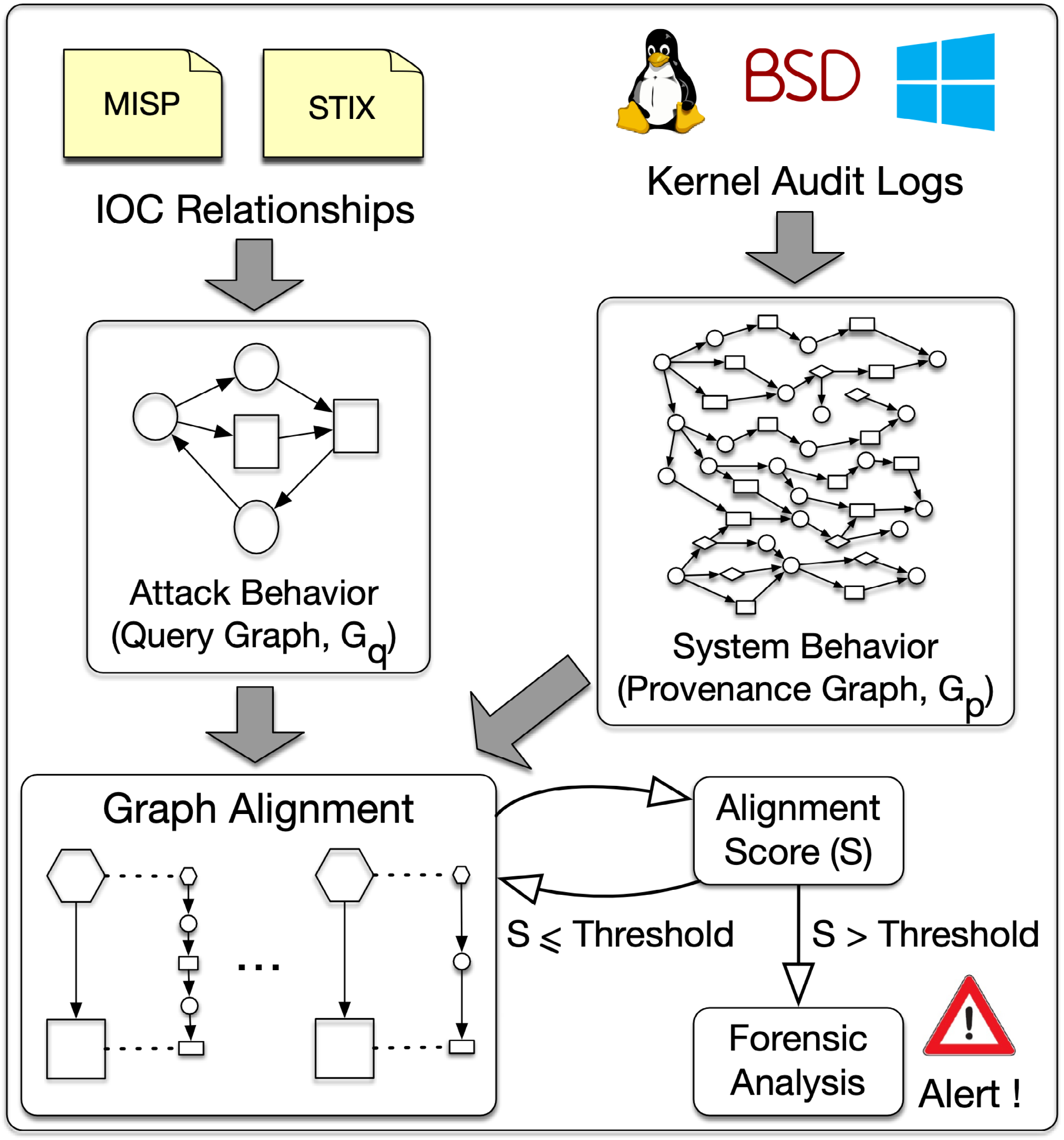}
 \vspace{1em}
 \caption{\projname Approach Overview.}
 \label{fig:approach_overview}
 \end{center}
 \end{figure}

A high-level view of our approach is shown in \Cref{fig:approach_overview}. We provide a brief overview of the components of \projname next, with more detailed discussions relegated to  \cref{sec:approach}.

\subsection{Provenance Graph Construction}\label{prov_graph_const}

To determine if the actions of the APT appear in the system, we model the kernel audit logs as a labeled, typed, and directed graph, which we call {\em provenance graph ($G_p$)}. This is a common representation of kernel audit logs, which allows tracking causality and information flow efficiently~\cite{king2003backtracking,king2005backtracking,gao2018aiql,gao2018saql,hossain2017sleuth}. In this graph, nodes represent system entities involved in the kernel audit logs, which have different types such as files and processes, while edges represent information flow and causality among those nodes taking into account the direction.
\projname currently supports consuming kernel audit logs\footnote{Kernel logs can be monitored using tools such as ETW, Auditd, and DTrace in Microsoft Windows, Linux, and FreeBSD, respectively.} from Microsoft Windows, Linux, and FreeBSD and constructs a provenance graph in memory, similar to prior work in this area \cite{hossain2017sleuth}. To support efficient searching on this graph, we leverage additional methods such as fast hashing techniques and reverse indexing for mapping process/file names to unique node IDs.

\subsection{Query Graph Construction}\label{query_graph_const}

We extract IOCs together with the relationships among them from CTI reports
related to a known attack. These reports appear in security blogs, threat intelligence reports by industry, underground forums on cyber threats, and public and private threat intelligence feeds. In addition to natural language, the attacks are often described in structured and semi-structured standard formats as well. These formats include OpenIOC\cite{open-ioc}, STIX\cite{stix}, MISP\cite{misp}, etc. Essentially, these exchange formats are used to describe the salient points of the attacks, the observed IOCs, and the relationships among them. For instance, using OpenIOC the behavior of a malware sample can be described as a list of artifacts such as the files it opens, and the DLLs it loads
~\cite{open-ioc-use-case}. 
These standard descriptions are usually created by the security operators manually \cite{stix1,stix2}. 
Additionally, automated tools have also been built to automatically extract IOCs from natural language and complement the work of human operators~\cite{zhu2018chainsmith,iace,ttpdrill}. These tools can be used to perform an initial extraction of features to generate the query graph and later refined manually by a security expert. We believe that manual refinement is an important component of the query graph construction because automated methods may often generate noise and reduce the quality of the query graphs. 

We model the behavior appearing in CTI reports also as a labeled, typed, and directed graph, which we call {\em query graph ($G_q$)}. If a description in a standard format is present, the creation of the query graph can be easily automated and further refined by humans. In particular, the entities appearing in the reports (e.g., processes, files) are transformed into nodes while relationships are transformed into directed edges \cite{stix-visulaization}. Nodes and edges of the query graph may be further associated with additional information such as labels (or names), types (e.g., processes, files, sockets, pipes, etc) and other annotations (e.g., hash values, creation time, etc) depending on the information that an analyst may deem necessary for matching. In the current \projname implementation, we use names and types
for specifying explicit mappings between nodes in the query graph and nodes in the provenance graph.

\begin{figure}[t]
  \begin{center}
    \includegraphics[width=.9\columnwidth]{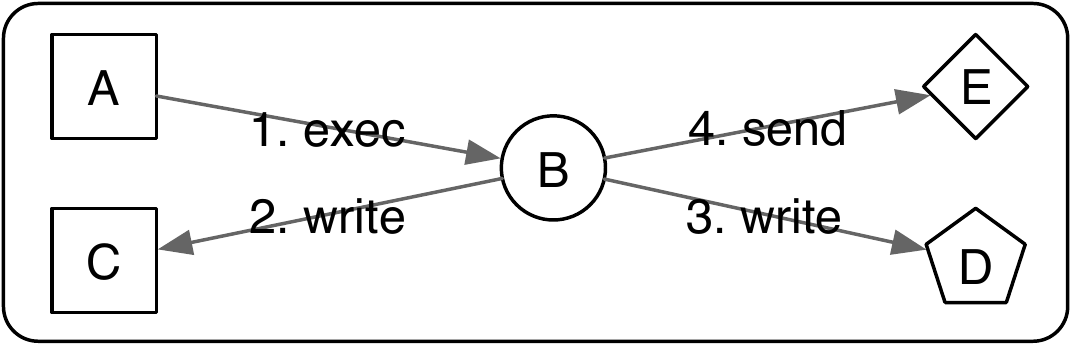}
  \end{center}
  \vspace{1em}
    \caption{Query Graph of DeputyDog Malware.
    A=$*$.\%exe\%, B=$*$, C=\%APPDATA\%\textbackslash $*$, D=\%HKCU\%\textbackslash Software\textbackslash Microsoft\textbackslash Windows\textbackslash CurrentVersion\textbackslash\- Run\textbackslash $*$, E=\%External\nespace IP\nespace address\%.
    }
    \label{fig:deputydog}
\end{figure} 

As an example of query graph construction, consider the following excerpt from a report \cite{deputydog-report} about the DeputyDog malware, used in our evaluation.\vspace{.5em}\\\vspace{.5em}
\fbox {
    \parbox{.95\columnwidth}{
\small{
Upon execution, 8aba4b5184072f2a50cbc5ecfe326701 writes ``28542CC0.dll'' to this location:
``C:\textbackslash Documents and Settings\textbackslash All Users\textbackslash Application Data\textbackslash 28542CC0.dll''.
In order to maintain persistence, the original malware adds this registry key:
``\%HKCU\%\textbackslash Software\textbackslash Microsoft\textbackslash Windows\textbackslash CurrentVersion\textbackslash Run\textbackslash\- 28542CC0''.
The malware then connects to a host in South Korea (180.150.228.102).}
}
}

The excerpt mentions several actions and entities that perform them and is readily transformed into a graph by a security analyst. For instance, the first sentence clearly denotes a process writing to a file (upon execution the malware writes a file to a location). We point out that the level of detail present in this excerpt is common across a large majority of CTI reports and can be converted to a reliable query graph by a qualified cyber analyst. In particular, the  verbs that express actions carried out by subjects can often be  easily mapped to reads/writes from/to disk or network and to interactions among processes (e.g., a browser downloads a file, a process spawns another process, a user clicks on a spear-phishing link, etc).

\Cref{fig:deputydog} shows the query graph corresponding to the above excerpt. 
Ovals, diamonds, rectangles, and pentagons represent processes, sockets, files, and registry entries, respectively \footnote{We use the same notation for the rest of the figures in the paper.}.
In \Cref{fig:deputydog}, node B represents the malware process or group of processes (we use a $*$ to denote that it can have any name), node A represents the image file of the malware, while nodes C, D and E represent a dropped file, a registry and an Internet location, respectively. We highlight at this point that the query graph that is built contains only information about information flows among specific entities as they appear in the report (processes, files, IP addresses, etc) and is not intended to be a precise subgraph of all the malicious entities that actually appear during the attack. In a certain sense, the query graph is a {\em summary} of the actual attack graph. 
 In our experiments, the  query graphs we obtained were usually small, containing between 10-40 nodes and up to 150 edges.

\subsection{Graph Alignment}
Finally, we model threat hunting as 
determining whether the query graph $G_q$ for the attack  ``manifests'' itself inside the provenance graph $G_p$. We call this problem {\em Graph Alignment Problem}.

We note at this point that $G_q$ expresses several high-level flows  between the entities (processes to files, etc.). In contrast, $G_p$ expresses the complete low-level activity of the system. As a result, an edge in $G_q$ might correspond to a path in $G_p$ consisting of multiple edges.  For instance,  if $G_q$ represents a compromised browser writing to a system file, in $G_p$ this may correspond to a path where a node representing a Firefox process forks new processes, only one of which ultimately writes to the system file. Often, this kind of correspondence may be created by attackers adding noise to their activities to escape detection. Therefore, we need a graph alignment technique that can {\em match single edges in $G_q$ to paths in $G_p$}. This requirement is critical in the design of our algorithm. 

In graph theory literature, there exist several versions of the graph matching problem. In {\em exact matching}, the subgraph embedded in a larger graph $G_p$ must be isomorphic to $G_q$ \cite{zong2014cloud}.
In contrast, in the {\em graph pattern matching} (GPM) problem,
some of the restrictions of exact matching are relaxed to extract more useful subgraphs.
However, both problems are NP-complete in the general case \cite{de2009subgraph}.
Even though a substantial body of work dedicated to \textit{GPM} exists \cite{gallagher2006matching,fan2010graph,cheng2008fast,zou2009distance,khan2013nema,gray,pienta2014mage}, many have limitations that make them impractical to be deployed in the field of {\em threat hunting}.
Specifically, they (i) are not designed for directed graphs with labels and types assigned to each node, (ii) do not scale to millions of nodes, or  (iii) are designed to  align all nodes or edges in the query graph exhaustively. 
Moreover, these approaches are not intended for the context of threat hunting, taking into account an evasive adversary which tries to remain stealthy utilizing the knowledge of the underlying matching criteria.
Due to these considerations,
we devise a novel graph pattern matching technique that addresses these limitations.
In  \Cref{fig:approach_overview}, graph nodes are represented in different shapes to model different node types, such as a file, process, and socket, however, the labels are omitted for brevity.
In particular, \projname starts by finding the set of all possible candidate alignments $i:j$ where  $i$ and $j$ represent nodes in $V(G_q)$ and $V(G_p)$, respectively. 
Then, starting from the alignment with the highest likelihood of finding a  match, called a {\em seed node},
we expand the search to find further node alignments.
The seed nodes are  represented by hexagons in \Cref{fig:approach_overview} while matching nodes in the two graphs are connected by dotted lines. 
To find an alignment that corresponds to the attack represented in CTI relationships, the search is expanded along paths that are more likely to be under the influence of an attacker. To estimate this likelihood, we devise a novel metric named {\em influence score}. Using this metric allows us to largely exclude irrelevant paths from the search and efficiently mitigate the {\em dependency explosion} problem.
Prior works have also proposed approaches to prioritize flows based on a {\em score} computed as length \cite{khan2013nema,gray} or cost \cite{hossain2017sleuth}.
However, they can be defeated  by attacks \cite{mimicry,asiaccs08} in which attackers frequently change their ways to evade the detection techniques. For instance, a proximity-based graph matching approach~\cite{khan2013nema,gray} might be easily evaded by attackers, who, being aware of the underlying system and matching approach, might generate a long chain of fork commands to affect the precision of proximity-based graph matching. In contrast, our score definition explicitly takes the influence of a potential attacker into account. 
In particular, we increase the cost for the attacker to evade our detection, by prioritizing flows based on the effort it takes for an attacker to produce them. Our search for alignment uses such prioritized flows and is described in \cref{sec:approach}.

After finding an alignment $G_q::G_p$, a score is calculated, representing the similarity between $G_q$ and the aligned subgraph of $G_p$.
When the score is higher than a threshold value, \projname raises an alert which declares the occurrence of an attack and presents a report of aligned nodes to a system analyst for further forensic analysis.
Otherwise, \projname  starts an alignment from the next seed node candidate.
After finding an attack subgraph in $G_p$, \projname generates a report containing the aligned nodes, information flows between them, and the corresponding timestamps.
In an enterprise setting, such visually compact and semantic-rich reports provide actionable intelligence to cyber analysts to plan and execute cyber-threat responses. We discuss the details of our approach in \cref{sec:approach}.

\section{ Algorithms}\label{sec:approach}

In this section, we discuss our main approach for alignment between $G_q$ and $G_p$ by (a) defining an {\em alignment metric} to measure how proper a graph alignment is, and (b) designing a best-effort similarity search based on specific domain characteristics.

\subsection{Alignment Metric}\label{sec:score_metric} 

\begin{table}[b]
\footnotesize

 \begin{center}
   \begin{tabular}{|M{1.2cm}|>{\raggedright\arraybackslash}M{6.2cm}|}
      \hline
      \textbf{Notation} & \textbf{Description} \\
        \hline
       $i:k$ & {\bf Node alignment}. Node $i$ is aligned to node $k$ ($i$ and $k$ are in two distinct graphs). \\
       \hline
       $i \xdashrightarrow{}j$ & {\bf Flow}. A path starting at node $i$ and ending at node $j$.\\
       \hline
       $i \xrightarrow[]{\text{label}} j$ & An edge from node $i$ to node $j$ with a specific label. \\
      \hline
       $G_q::G_p$ & {\bf Graph alignment}. A set of node alignments $i:k$ where $i$ is a node of $G_q$ and $k$ is a node of $G_p$.\\
        \hline
       $V(G)$ & Set of all vertices in graph $G$. \\
        \hline
       $E(G)$ & Set of all edges in graph $G$. \\
        \hline
       $F(G)$ & Set of all flows $i \xdashrightarrow{}j$ in graph $G$ such that $i \neq j$. \\
       \hline
   \end{tabular}
 \end{center}
\normalsize
\vspace{1em}
  \caption{Notations.}\label{notations}
\end{table}

\begin{figure*}[t]
\begin{center}
\includegraphics[width=\textwidth]{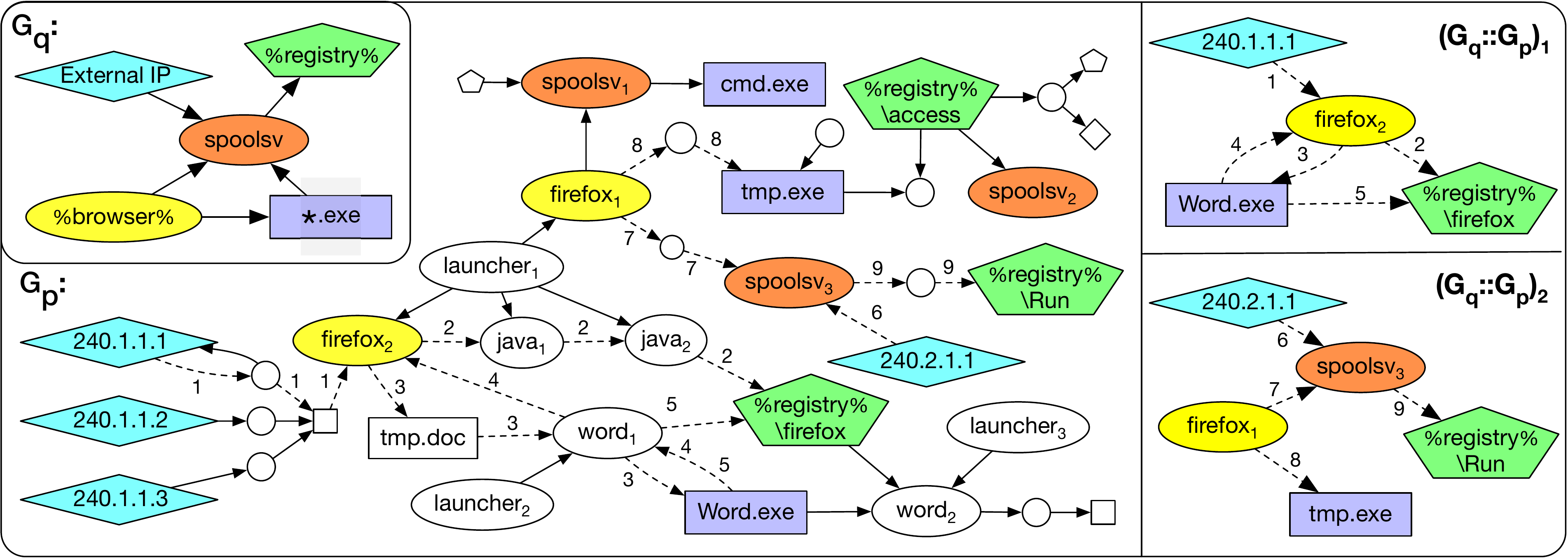}
\vspace{.5em}
\caption{Simplified Provenance Graph ($G_p$), Query Graph ($G_q$), and two sample graph alignments ($G_q::G_p$). Node types are shown with different shapes, and possible alignments for each node is shown with the same color. The numbers on the edges are merely to illustrate possible paths/flows and do not have additional meaning.}
\label{fig:sample_prov}
\end{center}
\end{figure*}

We introduce some notations (in \cref{notations}), where we define two kinds of alignments, i.e., a {\em node alignment} between two nodes in two different graphs, and a {\em graph alignment} which is a set of node alignments. Typically, two nodes $i$ and $j$ are in a \textit{node alignment} when they represent the same entity, e.g., a node representing a commonly used browser mentioned in the CTI report 
(node \textit{\%browser\%} in the query graph $G_q$ of \Cref{fig:sample_prov})
and a node representing a \textit{Firefox} process in the provenance graph. We note that, in general, the node alignment relationship is a many-to-many relationship from
V($G_q$) to V($G_p$), where V($G_q$) and V($G_p$) are the set of vertices of $G_q$ and $G_p$ respectively.
Therefore, given a query graph $G_q$, there may be a large number of \textit{graph alignments} between $G_q$ and many subgraphs of $G_p$. Another thing to point out is that each of these {\em graph alignments} can correspond to different subgraphs of $G_p$. Each of these subgraphs contains the nodes that are aligned with the nodes of $G_q$; however, they may contain different paths among those nodes. Among these subgraphs, we are interested in finding the subgraph that best matches the graph $G_q$.

Based on these definitions, the problem is to find the best possible graph alignment among a set of candidate graph alignments. To illustrate this problem, consider the query and provenance graphs $G_q$ and $G_p$, and two possible aligned graphs in \Cref{fig:sample_prov}, where the node shapes represent entity types (e.g., process, file, socket), and the edges represent information flow (e.g., read, write, IPC) and causal dependencies (e.g., fork, clone) between nodes.  The numbers shown on the edges of $G_p$ are not part of the provenance graph but serve to identify a single path in our discussion.
In addition, the subgraphs of $G_p$ determined by these two graph alignments with $G_q$ are represented by dotted edges in $G_p$. Each flow in $G_p$ and corresponding edge in $G_q$ is labeled with the same number. 
The problem is, therefore, to decide which among many alignments is the best candidate. Intuitively, for this particular figure, alignment $(G_q::G_p)_2$ is closer to $G_q$ than $(G_q::G_p)_1$, mainly because the number of its aligned nodes is higher than that of $(G_q::G_p)_1$, and most importantly, its flows have a better correspondence to the edges of the query graph $G_q$.

\subsubsection{Influence Score}\label{sec:influence_score}
Before formalizing the intuition expressed above, we must introduce a path scoring function, which we call {\em influence score} and which assigns a number to a given flow between two nodes. This score will be instrumental in defining the ``goodness'' of a graph alignment. In practice, the {\em influence score} represents the likelihood that an attacker 
can produce a flow.
To illustrate this notion, consider the two nodes \textit{firefox$_2$} and \textit{\%registry\%\textbackslash firefox} in the graph $G_{p}$ in \Cref{fig:sample_prov}. There exist two flows from \textit{firefox$_2$} to \textit{\%registry\%\textbackslash firefox}, one represented by the edges labeled with the number 2 (and passing through nodes java$_1$ and java$_2$), and another represented by the edges labeled 3, 3, and 5 (and passing through nodes tmp.doc and word$_1$). %
Assuming \textit{firefox$_2$} is under the control of an attacker, it is more likely for the attacker to execute the first flow rather than the second flow. In fact, in order to exercise the second flow, an attacker would have to take control over process launcher$_2$ or word$_1$ in addition to \textit{firefox$_2$}. Since launcher$_2$ or word$_1$ share no 
common ancestors in the process tree
with \textit{firefox$_2$}, such takeover would have to involve an additional exploit for launcher$_2$ or word$_1$, which is far more unlikely than simply exercising the first flow, where all processes share a common ancestor launcher$_1$.
We point out that this likelihood does not depend on the length of the flow, rather on the number of processes in that flow and on the number of distinct 
ancestors
those processes share in the process tree. One can, in fact, imagine a long chain of forked processes, which are however all under the control of the attacker because they all share a common 
ancestor
in the process tree, i.e., the first process of the chain. Another possible scenario of attacks present in the wild involves remote code loading from a single compromised process, where all the code with malicious functionality is loaded in main memory and the same process (e.g., firefox) executes all the actions on behalf of the attacker. While this technique leaves no traces on the file system and may evade some detection tools, \projname would be able to detect this kind of attack. In fact the influence score remains trivially unchanged. 

One additional important point to note is that this notion of measuring the potential {\em influence} of an attacker is very robust concerning evasion methods from an attacker. Every activity that an attacker may use to add noise and try to evade detection will likely have the same common 
ancestors, namely the initial compromise points of the attack, unless the attacker pays a higher cost to perform more distinct compromises.
Thus, such efforts will be ineffective in changing the  influence score of the paths.

Based on these observations, we define the {\em influence score}, $\Gamma_{i,j}$, between a node $i$ and a node $j$ as follows:

\small
\begin{equation}
\Gamma_{i,j}= \begin{cases}
\max\limits_{i\xdashrightarrow{}j}{\dfrac{1}{C_{min}(i\xdashrightarrow{}j)}} & \exists i\xdashrightarrow{}j \ | \ C_{min} (i\xdashrightarrow{}j) \le C_{thr}\\
0 & \text{otherwise}
\end{cases}
\label{eq:reachability_score}
\end{equation}
\normalsize

In \Cref{eq:reachability_score}, $C_{min}(i\xdashrightarrow{}j)$
represents the minimum number of distinct, independent compromises an attacker has to conduct to be able to generate the flow $i\xdashrightarrow{}j$.
This value captures the extent of the attacker's control over the flow and is calculated based on the minimum number of common 
ancestors
of the processes present in the flow.
For instance, if there is a flow from a node $i$ to a node $j$, and all the processes involved in that flow have one common 
ancestor in the process tree,
an attacker needs to compromise only that common 
ancestor 
process to initiate such flow, and therefore $C_{min}(i\xdashrightarrow{}j)$ is equal to 1. 
Note that if a node $i$ represents a process and  node $j$ a child process of $i$, then $C_{min}(i\xdashrightarrow{}j)$  will be equal to 1 as $i$ is parent of $j$.
If the number of common ancestors is larger than one (e.g., there are two ancestors in the path \textit{firefox$_2$} $ \rightarrow tmp.doc \rightarrow word_1 \rightarrow$ \textit{\%registry\%\textbackslash firefox}), an attacker has to compromise at least  as many (unrelated) processes independently; therefore it is harder for the attacker to construct such flow. For instance, for the attacker to control the flow \textit{firefox$_2$} $ \rightarrow tmp.doc \rightarrow word$\_1$ \rightarrow$ \textit{\%registry\%\textbackslash firefox}), (s)he needs to control both launcher$_1$ and launcher$_2$; therefore $C_{min}$ is equal to 2.

We also reasonably assume that it is not practical for an attacker to compromise more than a small number of processes with distinct exploits. In a vast majority of documented APTs, there is usually a single entry point or a very small number of entry points into a system for an attacker, e.g., a spear phishing email, or a drive-by download attack on the browser. We have confirmed that this is true also based on a review of a large number of white papers on APTs~\cite{aptnotes}. Once an attacker has an initial compromise, it is highly unlikely that they will invest additional resources in discovering and exploiting extra entry points.
Therefore, we can place a limit
$C_{thr}$
on the $C_{min}(i\xdashrightarrow{}j)$ values and reasonably assume that any flow between two nodes that has a $C_{min}(i\xdashrightarrow{}j)$ greater than $C_{thr}$ can not have been initiated by an attacker. 

While the value of $C_{min}(i\xdashrightarrow{}j)$ expresses how hard it is for an attacker to control a specific path, the {\em influence score} expresses how easy it is for an attacker to control that path, and it is defined as a value that is inversely proportional to $C_{min}(i\xdashrightarrow{}j)$. 
If there is more than one flow between two nodes $i$ and $j$, the {\em influence score} will be the maximum $\dfrac{1}{C_{min}(i\xdashrightarrow{}j)}$ over all those flows.
Based on this equation, the value of $\Gamma_{i,j}$ is  maximal (equal to 1) when there is a flow whose  $C_{min}(i\xdashrightarrow{}j)$ equals 1 and is minimal (equal to 0) when there is no flow with a $C_{min}(i\xdashrightarrow{}j)$ greater than $C_{thr}$.

\subsubsection{Alignment Score}
We are now ready to define a metric that specifies the score of a graph alignment $G_q::G_{p}$. 
Based on the notion of {\em influence score}, we define the scoring function $S(G_q::G_p)$, representing the score for an alignment $G_q::G_p$ as follows:

\small
\begin{equation}
S(G_q::G_p)=\frac{1}{|F(G_q)|}\sum_{ (i \xdashrightarrow{}j) \in F(G_q)} \Gamma_{k , l} \ |\  i:k \ \&\  j:l
\label{eq:new-scoring}
\end{equation}
\normalsize
In \Cref{eq:new-scoring}, nodes $i$ and $j$ are members of $V(G_q)$, and nodes $k$ and $l$ are members of $V(G_p)$. The flow $i \xdashrightarrow{}j$ is a flow defined over $G_q$. In particular, the formula starts by computing the sum of the influence scores among all the pairs of nodes $(k, l)$ with at least one path from $k$ to $l$ in the graph $G_p$ such that $k$ is aligned with $i$ and $l$ is aligned with $j$.  This sum is next normalized by dividing it with the maximal value possible for that sum. In fact, $|F(G_q)|$ is the number of flows in $G_q$. Since the maximal value of the {\em influence score} between two nodes is equal to 1, then the number of flows automatically represents the maximal value of the sum of the {\em influence scores}.

From \Cref{eq:new-scoring}, intuitively, the larger the value of $S(G_q::G_p)$, the larger the number of node alignments and the larger the similarity between flows in $G_q$ and flows in $G_p$, which are likely to be under the influence of a potential attacker. In particular, the value of $S(G_q::G_p)$  is between 0 and 1. When $S(G_q::G_p)=0$, either no proper alignment is found for nodes in $V(G_q)$, or no similar flows to those of $G_q$ appear between the aligned nodes in $G_p$.
On the contrary, when $S(G_q::G_p)=1$, all the nodes in $G_q$ are aligned to the nodes in $G_p$, and all the flows existing in $G_q$ also appear between the aligned nodes in $G_p$, and they all have an {\em influence score} equal to $1$, i.e., it is highly likely that they are under the attacker's control. 

Finally, when the alignment score $S(G_q::G_p)$ bypasses a predetermined threshold value ($\tau$), we raise the alarm. To determine the optimal value of this threshold, recall that $C_{thr}$ is the maximum number of distinct entry point processes we are assuming an attacker is willing to exploit independently.
Therefore, an attacker is assumed to be able to influence any information flow with {\em influence score} of $\frac{1}{C_{thr}}$ or higher.
On the other hand, $S(G_q::G_p)$ is the average of all {\em influence scores}. 
Therefore, we define the threshold $\tau$ as follows:
\small
\begin{align}
S(G_q::G_p) \ge\tau \\
\tau = \frac{1}{C_{thr}}
\label{eq:threshold}
\end{align}
\normalsize
If $S(G_q::G_p)$ bypasses $\tau$, we declare a match and raise the alarm.

\subsection{Best-Effort Similarity Search}\label{best-effort}
After defining the alignment score, we describe our procedure to search for an alignment that maximizes that score. In particular, given a query graph $G_q$, we need to search a very large provenance graph $G_p$ to find an alignment $G_q::G_p$ with the highest alignment score based on \Cref{eq:new-scoring}.

The first challenge in doing this is the size of $G_p$, which can reach millions of nodes and edges. Therefore, it is not practical to store {\em influence scores} between all pairs of nodes of $G_p$. We need to perform graph traversals on demand to find the {\em influence scores} between nodes or even to find out whether there is a path between two nodes.
Besides, we are assuming that all analytics are being done on a stationary snapshot of $G_p$, and no changes happen to its nodes or edges from the moment when a search is initiated until it terminates.

Our search algorithm consists of the following four steps, where steps 2-4 are repeated until finding alignment with a score higher than the threshold value $\tau$ (\Cref{eq:threshold}).

{\bf Step 1. Find all Candidate Node Alignments}:  
We start by searching among nodes of $G_p$ to find candidate alignments for each node in $G_q$.
These candidate alignments are chosen based on the name, type, and annotations on the nodes of the query graph.
For instance, nodes of the same type (e.g., two process nodes) with the same label (e.g., Firefox) appearing in $G_q$ and $G_p$ may form candidate alignments, nodes whose labels match a regular expression (e.g., a file system path and file name), and so on. A user may also manually annotate a node in the provenance graph and explicitly specify an alignment with a node in the query graph. In general, a node in $G_q$ may have any number of possible alignments in $G_p$, including 0.
Note that in this first step, we do not have enough information about paths and flows and are looking at nodes in isolation. In \Cref{fig:sample_prov}, the candidate node alignments are represented by the pairs of nodes having the same color.

{\bf Step 2. Selecting Seed Nodes}:  
To find a good-enough alignment $G_q::G_p$, we need to explore connections between candidate alignments found in Step 1, by performing graph traversals on $G_p$.
However, due to the structure and large size of $G_p$, starting a set of graph traversals from randomly aligned nodes in  $G_p$ might lead to costly and unfruitful searches. To determine a {\em good} starting point, a key observation is that the attack activities usually comprise  a tiny portion of $G_p$, while benign activities are usually repeated multiple times. Therefore, it is more likely for artifacts that are specific to an attack to have fewer alignments than artifacts of benign activities. Based on this observation, we sort the nodes of $G_q$ by an increasing order in the number of candidate alignments related to each node.  We select the seed nodes with fewest alignments first. For instance, with respect to the example in  \Cref{fig:sample_prov}, the seed node will be \textit{\%browser\%}, since it has the smallest number of candidate node alignments. 
If there are seed nodes with the same number of candidate alignments, we choose one of them randomly. 

{\bf Step 3. Expanding the Search}: 
In this step, starting from the seed node chosen at Step 2, we iterate over all the nodes in $G_p$ aligned to it and initiate a set of graph traversals, going forward or backward, to find out whether we can reach other aligned nodes among those found in Step 1. For instance, after choosing node \textit{\%browser\%} as a seed node, we start a series of forward and backward graph traversals from the nodes in $G_p$ aligned to \textit{\%browser\%}, that is \textit{firefox$_1$} and \textit{firefox$_2$}. In theory, these graph traversals can be very costly both because of the size of the graph and also the number of candidate aligned nodes, which can be located anywhere in the graph. In practice, however, we can stop expanding the search along a path once the {\em influence score} between the seed node and the last node in that path reaches 0. 
For instance, suppose we decide that $C_{thr}$ is equal to 2 in  \Cref{fig:sample_prov}. Then, the search along the path (\textit{firefox$_2$} $ \rightarrow tmp.doc \rightarrow word_1 \rightarrow $ \textit{\%registry\%\textbackslash firefox}$ \rightarrow word_2$) will not expand past the node word$_2$, since the $C_{min}$ between \textit{firefox$_2$} and any node along that path becomes greater than 2 at word$_2$, and thus the influence score becomes $0$. Note that there is an additional path from \textit{firefox$_2$} to word$_2$ via \textit{\%registry\%\textbackslash firefox} and  along this path, the $C_{min}$ between \textit{firefox$_2$} and word$_2$ is still 2. Therefore, because of this path, the search will continue past word$_2$. Using the {\em influence score} as an upper bound in the graph traversals dramatically reduces the search complexity and enables a fast exploration of the graph $G_p$.

Based on the shape of the query graph $G_q$, multiple forward/backward tracking cycles might be required to visit all nodes (for instance, if we choose \textit{\%browser\%} as a seed node in our example, then node \textit{240.2.1.1} in $G_p$ is unreachable with only one forward or backward traversal starting at \textit{firefox$_1$} or \textit{firefox$_2$}). In this case, we repeat the backward and forward traversals starting from nodes that are adjacent to the unvisited nodes but that have been visited in a previous traversal (for instance, node \textit{spoolsv$_3$} in our example). We iterate this process until we cover all the nodes of the query graph $G_q$. 
{\bf Step 4. Graph Alignment Selection}: 
This step is responsible for producing the final result or for starting another iteration of the search from step 2, in case a result is not found. In particular, after performing backward/forward traversals, we identify a subset of candidate nodes in $G_p$ for each node in $G_q$. For instance, with respect to our example, we find that node\textit{\%browser\%} has candidates \textit{firefox$_1$} and \textit{firefox$_2$}, node \textit{External IP} has candidate alignments \textit{240.1.1.1}, \textit{240.1.1.2}, \textit{240.1.1.3}, and \textit{240.2.1.1}, and so on. However, the number of possible candidate {\em graph alignments} that these candidate nodes can form can be quite large. If each node $i$ in $G_q$ has $n_i$ candidate alignments, then the number of possible graph alignments is equal to ${\displaystyle \prod_{i} n_i}$. For instance, in our example, we can have 216 possible graph alignments ($2\times 3\times 3\times 3\times 4$). In this step, we search for the graph alignment that maximizes the alignment score (\Cref{eq:new-scoring}). 

A naive method for doing this is a brute-force approach that calculates the alignment score for all possible graph alignments. However, this method is very inefficient and does not fully take advantage of domain knowledge. To perform this search efficiently, we devise a procedure that iteratively chooses the best candidate for each node in $G_q$ based on an approximation function that measures the maximal contribution of each alignment to the final alignment score. 

In particular, starting from a seed node in $G_q$, we select the node in $G_p$ that maximizes the contribution to the alignment score and fix this node alignment (we discuss the selection function in the next paragraph). For instance, starting from seed node \textit{\%Browser\%} in our example, we fix the alignment with node \textit{firefox$_1$}. From this fixed node alignment, we follow the edges in $G_q$ to fix the alignment of additional nodes connected to the seed node. The specific node alignment selected for each of these nodes is the one that maximizes the contribution to the alignment score. For instance, from node \textit{\%Browser\%} (aligned to \textit{firefox$_1$}), we can proceed to node \textit{$*$.exe} and fix the alignment of that with one node among \textit{cmd.exe}, \textit{tmp.exe}, and \textit{Word.exe}, such that the contribution to the alignment score is maximized. 

\noindent
\textbf{Selection Function}. The key intuition behind the selection function, which selects and fixes one among many node alignments, is to approximate how much each alignment would contribute to the final alignment score and to choose the one with the highest contribution. For a given candidate aligned node $k$ in $G_p$, this contribution is calculated as the sum of the maximum influence scores between that node and all the other candidate nodes $l$ in $G_p$ that: 1) are reachable from $k$ or that have a path to $k$, and 2) whose corresponding aligned node $j$ in $G_q$ has a flow from/to the node in $G_q$ that corresponds to node $k$.  
For instance, consider node \textit{\%Browser\%} and the two candidate alignment nodes \textit{firefox$_1$} and \textit{firefox$_2$} in our example. To determine the contribution of \textit{firefox$_1$}, we measure for every flow (\textit{\%Browser\%} $\xdashrightarrow{}$ \textit{$*$.exe}, \textit{\%Browser\%} $\xdashrightarrow{}$ \textit{spoolsv}, \textit{\%Browser\%} $\xdashrightarrow{}$ \textit{\%registry\%}) from/to \textit{\%Browser\%} in $G_q$, the maximum {\em influence score} between \textit{firefox$_1$} and the candidate nodes aligned with \textit{$*$.exe}, \textit{spoolsv}, and \textit{\%registry\%}, respectively. In other words, we compute the maximum {\em influence score} between \textit{firefox$_1$} and each of the node alignment candidates of \textit{$*$.exe}, the maximum {\em influence score} between \textit{firefox$_1$} and each of the node alignment candidates of \textit{spoolsv}, and the maximum {\em influence score} between \textit{firefox$_1$} and each of the node alignment candidates of \textit{\%registry\%}. Each of these three maximums provides the maximal contribution to the alignment score of each of the possible future alignments (which are not fixed yet) for \textit{$*$.exe}, \textit{spoolsv}, and \textit{\%registry\%}, respectively. Next, we sum these three maximum values to obtain the maximal contribution that \textit{firefox$_1$} would provide to the alignment score. We repeat the same procedure for \textit{firefox$_2$} and, finally select the alignment with the highest contribution value.
This contribution is formally computed by the following equation, which approximates $A(i:k)$ the contribution of a node alignment $i:k$.

\small
 \begin{align}
 \begin{split}
&A(i:k)\\
&=\hspace{-1.5em} \sum_{j : (i \xdashrightarrow{}j) \in F(G_q)} \hspace{-.25em} \Big( 1_{\{j:l\}} \times \Gamma_{k , l} + (1- 1_{\{j:l\}}) \times \hspace{-0.5em} \max_{m \in candidates(j)} (\Gamma_{k , m}) \Big) \\
& +  \hspace{-1.5em} \sum_{j : (j \xdashrightarrow{}i) \in F(G_q)} \hspace{-.25em} \Big( 1_{\{j:l\}} \times \Gamma_{l , k} + (1-1_{\{j:l\}}) \times \hspace{-0.5em} \max_{m \in candidates(j)}(\Gamma_{m , k}) \Big) 
\label{eq:new-scoring-approx}
\end{split}
\end{align}
\normalsize

where $1_{\mathcal{A}}$ is an indicator function, which is $1$ if the alignment expressed in $\mathcal{A}$ is fixed, and is $0$ otherwise.
In other words, if the alignment between node $j$ and $l$, has been fixed, $l_{\{j:l\}}$ equals to $1$, and otherwise, if node $j$ is not aligned to any node yet, $1_{\{j:l\}}$  equals to $0$. Note that $1_{\{j:l\}}$ and $(1-1_{\{j:l\}})$ are mutually exclusive, and at any moment, only one of them equals $1$, and the other one equals to $0$.

We note that the first summation is performed on outgoing flows from node $i$, while the second summation is performed on flows that are incoming to node $i$.
Inside each summation, the first term represents a fixed alignment while the second term represents the maximum among potential alignments that have not been fixed yet, as discussed above.

Finally, for each node $i$ having a set $K$ of candidate alignments as produced by Step 3, the selection function, which fixes the alignment of $i$ is as follows:

\small
\begin{equation}
\argmax_{k\in K}{A(i:k)}
 \label{eq:selection}
\end{equation}
\normalsize

The intuition behind equations \ref{eq:new-scoring-approx} and \ref{eq:selection} is that once a node alignment is fixed, the other possible alignments of that node are ignored by future steps of the algorithm and the calculation of the maximum influence score related to that alignment is reduced to a table lookup instead of an iteration over candidate node alignments. In particular, the search starts as a brute force search, but as more and more node alignments are fixed, the search becomes faster by reusing results of previous searches stored in the table. Using equations \ref{eq:new-scoring-approx} and \ref{eq:selection} dramatically speeds up the determination of a proper graph alignment. While in theory, this represents a greedy approach, which may not always lead to the best results, in practice, we have found that it works very well.
Finally, after fixing all node alignments, the alignment score is calculated as in \Cref{eq:new-scoring}. If the score is below the threshold, the steps 2-4 are executed again.
Our evaluation results in \cref{sec:eval} show that the attack graph is usually found within the first few iterations.

\section{Evaluation}\label{sec:eval}

\begin{figure*}[t]
  \begin{center}

    \includegraphics[width=.65\columnwidth]{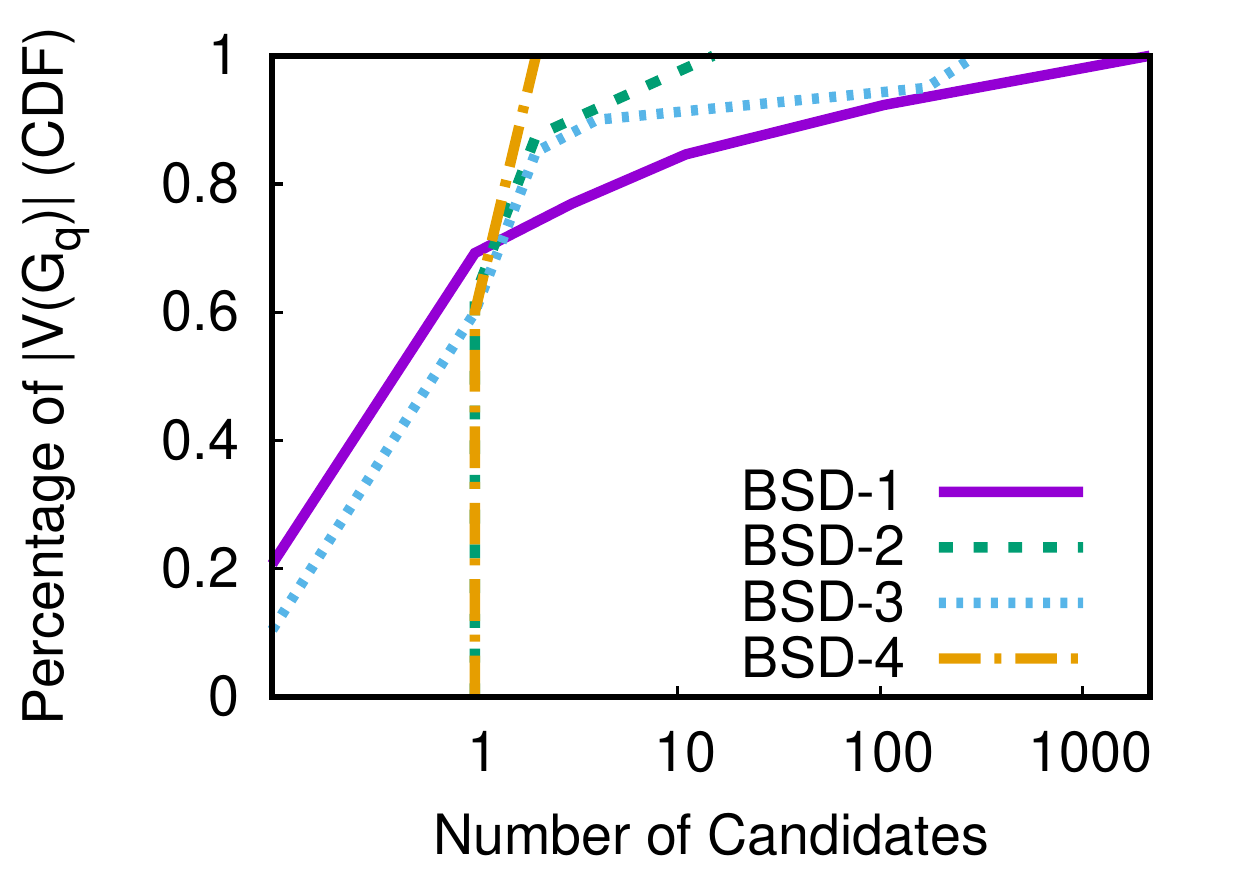}\hfill
    \includegraphics[width=.65\columnwidth]{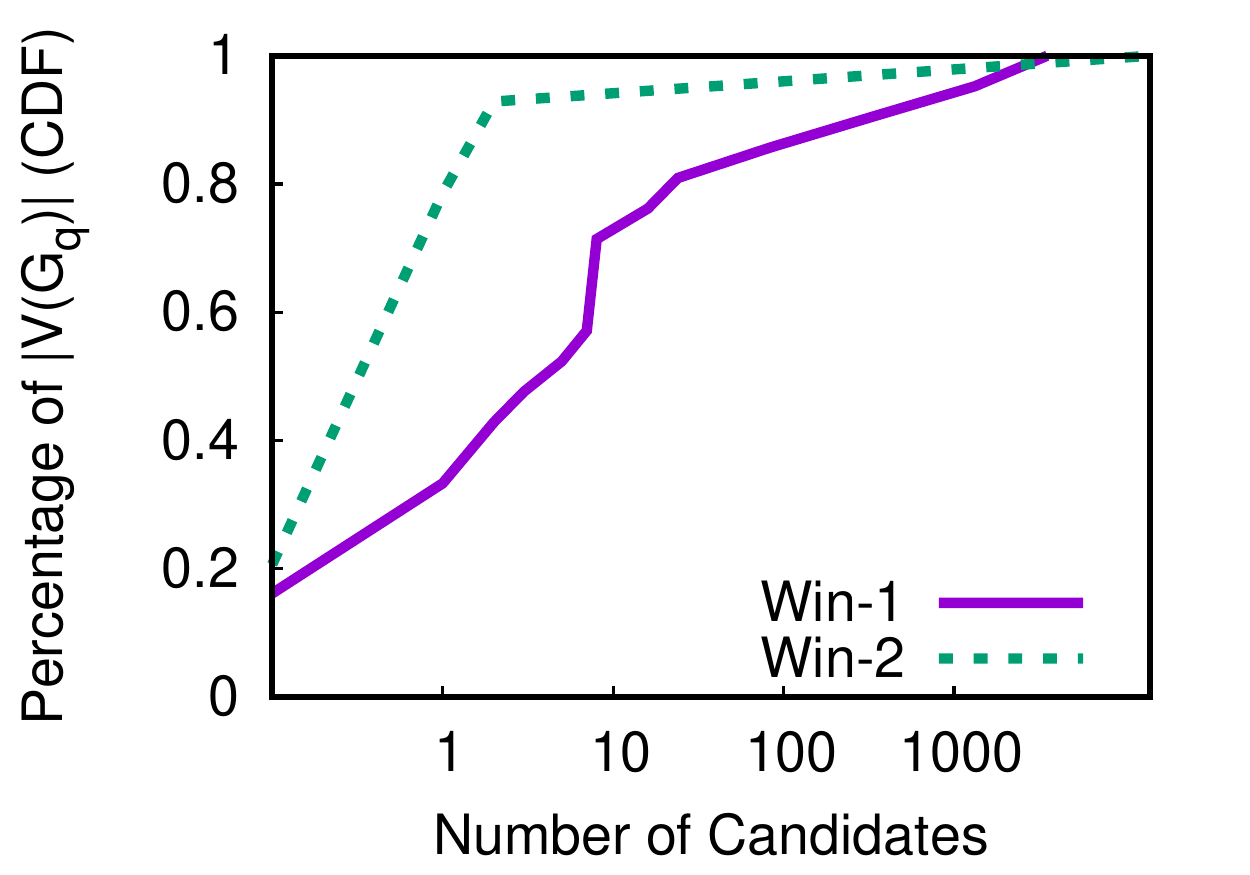}\hfill
    \includegraphics[width=.65\columnwidth]{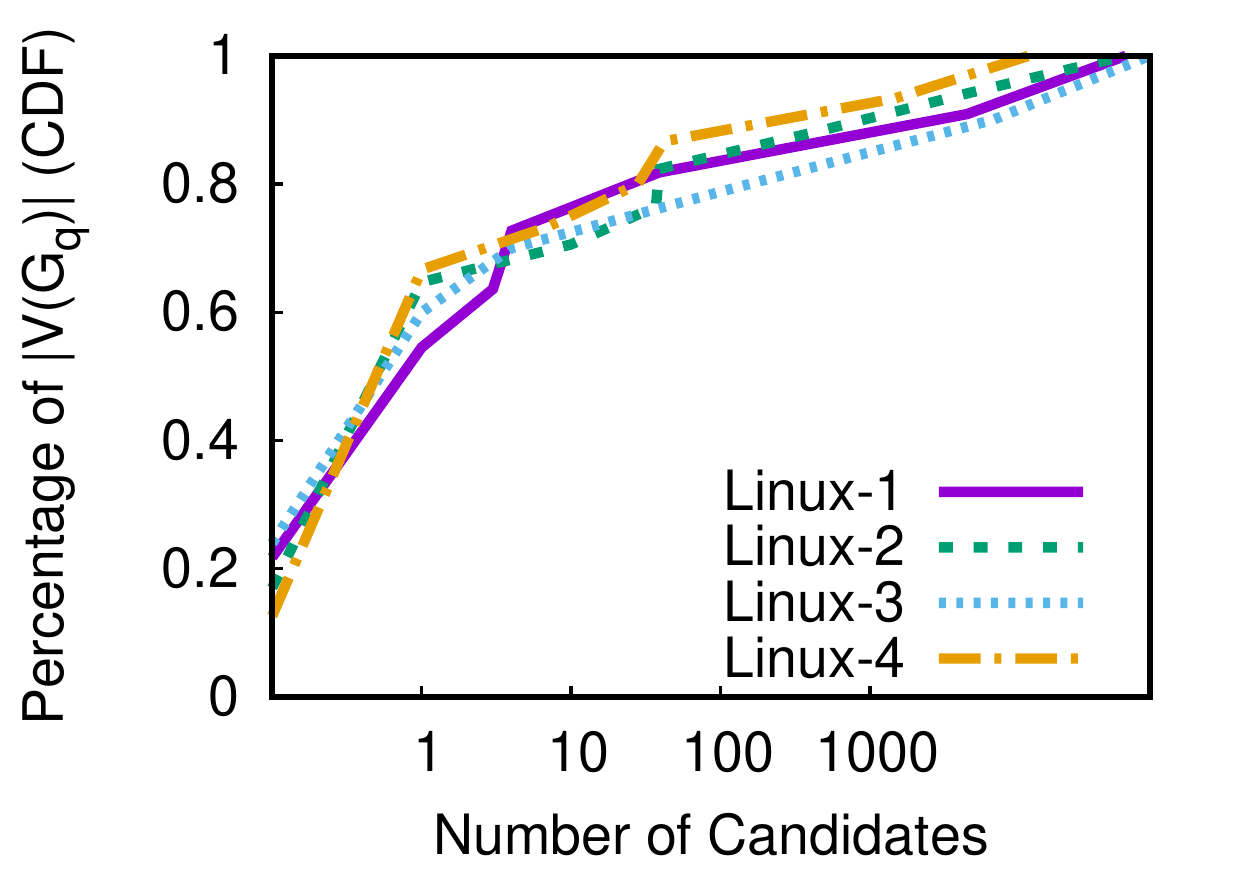}
  \end{center}
  \vspace{.5em}
    \caption{Cumulative Distribution Function (CDF) of number of candidates in $|G_p|$ for each node of $|G_q|$. From left to right: BSD, Windows, and Linux Scenarios.}\label{fig:candid}
\end{figure*}

We evaluate \projname's efficacy and reliability in three different experiments.
In the first experiment, we use a set of DARPA Transparent Computing (TC) program red-team vs. blue-team adversarial engagement scenarios which are set up in an isolated network simulating an enterprise network. The setup contains target hosts (Windows, BSD, and Linux) with kernel-audit reporting enabled.
During the engagement period, benign background activities were continuously run in parallel to the attacks from the red team.

In the second experiment, we further test \projname on real-world incidents whose natural language descriptions are publicly available on the internet. To reproduce the attacks described in the public threat reports, 
we obtained and executed their binary samples in a controlled environment and collected kernel audit logs from which we build the provenance graph.
In the third experiment, we evaluate \projname's robustness against false signals in an attack-free dataset.

In all the experiments, we set the value of $C_{thr}$ to $3$ (and thus a threshold of $\frac{1}{3}$).
This choice is 
validated in \cref{validating_threshold}. 
We note, however, that one can configure \projname  with higher or lower values depending on the 
confidence about the system's protection mechanisms or the effort cyber-analysts are willing to spend to check the alarms. In fact,
 the value of $C_{thr}$ influences the number of false positives and potential false negatives. A higher $C_{thr}$ will increase the number of false positives while a lower $C_{thr}$ will reduce it. On the other hand, a higher value of $C_{thr}$ may detect sophisticated attacks, with multiple initial entry points, while a smaller value may miss them. 
After finding alignment with a score bypassing the threshold, we manually analyzed all the matched attack subgraphs to confirm  that they were correctly pinpointing the actual attacks present in the query graphs.

\subsection{Evaluation on the DARPA TC Dataset}\label{eval:tc}

\begin{table}[b]
 \footnotesize
 \begin{center}
   \begin{tabular}{|M{1cm}|M{1.4cm}|M{1.4cm}|M{1cm}|M{1cm}|}
      \hline
      \textbf{Scenario} & \textbf{subjects $\in$ \textbar $V(G_q)$\textbar} & \textbf{objects $\in$ \textbar $V(G_q)$\textbar} & \textbf{\textbar $E(G_q)$\textbar} & \textbf{\textbar $F(G_q)$\textbar}\\
      \hline
      BSD-1  & 4 & 9 & 19 & 81 \\
      \hline
      BSD-2  & 1 & 7 & 10 & 32 \\
      \hline
      BSD-3  & 3 & 18 & 34 & 159 \\
      \hline
      BSD-4 & 2 & 8 & 13 & 43 \\
      \hline
      Win-1 & 13 & 8 & 26 & 149 \\
      \hline
      Win-2 & 1 & 13 & 19 & 94 \\
      \hline
      Linux-1 & 2 & 9 & 19 & 62 \\
      \hline
      Linux-2 & 5 & 12 & 24 & 112 \\
      \hline
      Linux-3 & 2 & 8 & 22 & 48 \\
      \hline
      Linux-4 & 4 & 11 & 22 & 96 \\
      \hline
                  
   \end{tabular}
 \end{center}
 \normalsize
\vspace{1em}
 \caption{Characteristics of Query Graphs.}\label{query_graph}
\end{table}

This experiment was conducted on a dataset released by the DARPA TC program, generated during a red-team vs. blue-team adversarial engagement in April 2018 \cite{tc_github}. 
In the engagement, different services were set up, including a web server, an SSH server, an email server, and an SMB server. An extensive amount of benign activities was simulated, including system administration tasks, web browsing to many web sites, downloading, compiling, and installing multiple tools.
The red-team relies on threat descriptions to execute these attacks. We  obtained these threat descriptions and used them to extract a query graph for each scenario (summary shown in \cref{query_graph}).
In total, we evaluated \projname on
ten attack scenarios including four on BSD, two on Windows, and four on Linux. 
Due to space restrictions, we are not able to show all the query graphs; however, their characteristics are described in \cref{query_graph}, where subjects indicate processes, and objects indicate files, memory objects, and sockets. 
BSD-1-4 pertain to attacks conducted on a FreeBSD 11.0 (64-bit) web-server which was running a back-doored version of Nginx. Win-1\&2 pertain to attacks conducted on a host machine running Windows 7 Pro (64-bit). The Win-1 scenario contains a phishing email with a malicious Excel macro attachment, while the Win-2 scenario contains exploitation of a vulnerable version of the Firefox browser. Linux1\&2 and Linux3\&4 pertain to attacks conducted on hosts running Ubuntu 12.04 (64-bit) and Ubuntu 14.04 (64-bit), respectively. Linux1\&3 contain in-memory browser exploits, while Linux2\&4 involve a user who is using a malicious browser extension.

\begin{figure*}[t]
  \begin{center}

    \includegraphics[width=\textwidth]{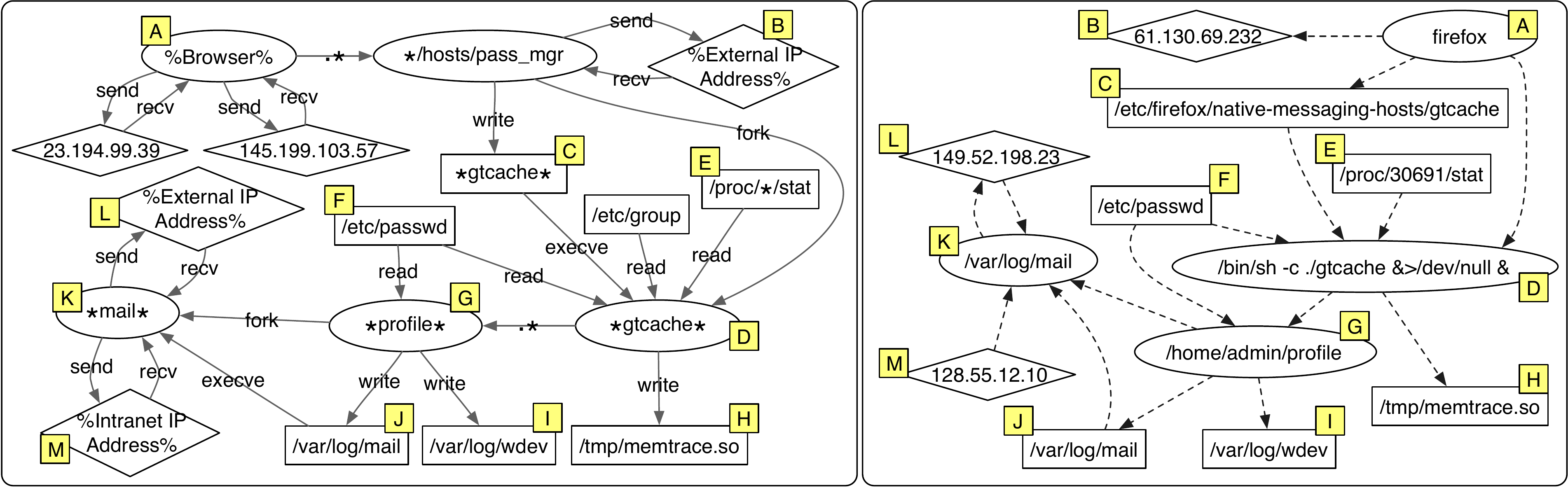}
  \end{center}
  \vspace{.5em}
    \caption{Query Graph of Scenario: Linux-2 (on the left) and its Detected Alignment (on the right).}
    \label{fig:linux2}
\end{figure*}

\begin{table}[b]
 \footnotesize
 \begin{center}
   \begin{tabular}{|M{1.1cm}|M{1.4cm}|M{1.3cm}|M{1.1cm}|M{1.4cm}|}
      \hline
      \textbf{Scenario} & \textbf{Earliest iteration bypassing threshold} & \textbf{Earliest score bypassing threshold} & \textbf{Max score in 20 iterations} & \textbf{Earliest iteration resulting Max score} \\
      \hline  
      BSD-1 & 1 & 0.45 & 0.64 & 5 \\
      \hline    
      BSD-2 & 1 & 0.81 & 0.81 & 1 \\
      \hline
      BSD-3 & 1 & 0.89 & 0.89 & 1 \\
      \hline
      BSD-4 & 1 & 1 & 1 & 1 \\
      \hline
      Win-1 & 1 & 0.63 & 0.63 & 1 \\
      \hline
      Win-2 & 1 & 0.47 & 0.63 & 4 \\
      \hline      
      Linux-1 & 1 & 0.58 & 0.58 & 1 \\
      \hline
      Linux-2 & 2 & 0.55 & 0.71 & 5 \\
      \hline
      Linux-3 & 1 & 0.54 & 0.54 & 1 \\
      \hline
      Linux-4 & 1 & 0.87 & 0.87 & 1 \\
      \hline
   \end{tabular}
 \end{center}
 \normalsize
 \vspace{1em}
 \caption{\projname's Graph Alignment Scores.
 }\label{TC_scores}

\end{table}

\noindent
\textbf{Alignment Score.}\label{sec:scores}
As discussed in \cref{best-effort}, \projname iteratively repeats the node alignment procedure starting from the seed nodes with fewer candidates.
\Cref{fig:candid} shows the number of candidate aligned nodes for each node of $G_q$. Most of the nodes of $G_q$ have less than ten candidate nodes in $G_p$, while there are also nodes with thousands of candidate nodes. These nodes, which appear thousands of times, are usually ubiquitous processes and files routinely accessed by benign activities, such as Firefox or Thunderbird. We remind the reader that our seed nodes are chosen first from the nodes with fewer alignments. In each iteration, an alignment is constructed, and its alignment score is compared with the threshold value, which is set to $\frac{1}{3}$. 

\Cref{TC_scores} shows \projname's matching results for each DARPA TC scenario after producing an alignment of the query graphs with the corresponding provenance graphs. We stop the search after the first alignment that surpasses the threshold value. 
The second and third columns of \cref{TC_scores} show the number of iterations of the steps 2-4 presented in  \cref{best-effort} and the actual score obtained for the first alignment that bypasses the threshold value. 
In 9 out of 10 scenarios, an alignment bypassing the threshold value was found in the first iteration.
In one case, the exact matching of $G_q$  could be found in $G_p$ (see BSD-4).

The fourth column of \cref{TC_scores} shows the maximum alignment score among the 20 alignments constructed by iterating steps 2-4 of our search algorithm 20 times while 
the last column shows the earliest iteration-number that resulted in the maximum value.
As can be seen, on average, our search converges quickly to a perfect solution. In 7 out of 10 scenarios, the maximum alignment score is calculated in the first iteration, while in the other 3, the maximum alignment scores are calculated in the fourth or fifth iterations.
 The latter is due to slight differences between the attack reports and the red team's implementation of the attacks, which result in information flows and causal dependencies that differ slightly between the query graph and the provenance graph.
As an example, in \Cref{fig:linux2}, we show the query graph and its aligned subgraph for the Linux-2 scenario. 
In this scenario, the attacker exploits Firefox via a malicious password manager browser extension, to implant an executable to disk. 
Then, the attacker runs the dropped executable to exfiltrate some confidential information and perform a port scan of known hosts on the target network.
We tag the aligned nodes in each graph with the same letter label. Some nodes on the query graph are not aligned with any nodes in the provenance graph. This reduces the score of the graph alignment to a value that is less than 1. 
Although  $G_q$ largely overlaps with a subgraph in $G_p$, some nodes have no alignment, and some information flows and causal dependencies do not appear in the provenance graph. The percentage of these nodes is small, however. As long as the reports are mainly matching the actual attack activities, our approach will not suffer from this.

\subsection{Evaluation on Public Attacks}\label{sec:eval:malware}
In this section, we describe the evaluation of \projname on attacks performed by real-world malware families and compare its effectiveness with that of other similar tools. We show the results of this evaluation in \cref{malware-db}.
The names of these malware families, the CTI reports we used as descriptions of their behavior, and the year in which the report is published are shown in the first three columns.

\begin{table*}[!ht]
 \footnotesize
 \begin{center}
    \begin{tabular}{|M{1.2cm}|M{1.5cm}|M{0.5cm}|M{1cm}|M{3.6cm}|M{.9cm}|M{0.8cm}|M{1cm}|M{0.7cm}|M{0.7cm}|M{1.4cm}|}
      \hline
      \textbf{Malware} & \textbf{Report} & \multirow{ 2}{*}{\textbf{Year}} & \textbf{Reported} & \multirow{ 2}{*}{\textbf{Analyzed Malware MD5}} & \textbf{Sample} & \textbf{Isolated} 
      & \multicolumn{4}{|c|}{\textbf{Detection Results}}\\
      \cline  {8-11}
      \textbf{Name} & \textbf{Source} &  & \textbf{Samples} &  & \textbf{Relation} & \textbf{IOCs} & 
      \textbf{RedLine} & \textbf{Loki} & \textbf{Splunk} & \textbf{\projname}\\
      \hline
      njRAT & Fidelis~\cite{njrat-report} & 2013 & 30 & 2013385034e5c8dfbbe47958fd821ca0 & different & 153 & 
      F+H & F+H & P & B (score=0.86)\\
      \hline    
      DeputyDog & FireEye~\cite{deputydog-report} & 2013 & 8 & 8aba4b5184072f2a50cbc5ecfe326701 & subset & 21 & 
      F$\times 2$+H+R & F$\times 2$+H & P+R & B (score=0.71)\\
      \hline
      Uroburos & Gdata~\cite{uroburos-report} & 2014 & 4 & 51e7e58a1e654b6e586fe36e10c67a73 & subset & 26 
      & F+H & F+H & R & B (score=0.76)\\
      \hline
      Carbanak & Kaspersky~\cite{carbanak-report} & 2015 & 109 & 1e47e12d11580e935878b0ed78d2294f & different & 230 & 
      - & PE & S & B (score=0.68)\\
      \hline
      DustySky & Clearsky~\cite{dustysky-report} & 2016 & 79 & 0756357497c2cd7f41ed6a6d4403b395 & different & 250 &
      - & - & - & B (score=1.00)\\
      \hline
      OceanLotus & Eset~\cite{oceanlotus-report} & 2018 & 9 & d592b06f9d112c8650091166c19ea05a & subset & 117 &
      F+R & F+PE & P+R & B (score=0.65)\\
      \hline
      HawkEye & Fortinet~\cite{hawkeye-report} & 2019 & 3 & 666a200148559e4a83fabb7a1bf655ac & different & 3 & 
      - & PE & - & B (score=0.62)\\
      \hline
    \end{tabular}
 \end{center}
  \normalsize
  \vspace*{1em}
  \caption{Malware reports. In the Detection Results, B=Behavior, PE=PE-Sieve, F=File Name, H=Hash, P=Process Name, R=Registry, S=Windows Security Event.}\label{malware-db}
  \vspace*{-.4em}
\end{table*}

\afterpage{
\begingroup
\let\clearpage\relax 

\begin{table*}
 \footnotesize
  \vspace*{-1em}
   \begin{tabular}{|M{1.1cm}|M{.5cm}|>{\raggedright\arraybackslash}M{7.3cm}|>{\raggedright\arraybackslash}M{7.3cm}|}
      \hline
      \textbf{Malware} & \textbf{Node} & \textbf{Label} & \textbf{Aligned Node Label}\\
      \hline  
      \multirow{10}{*}{Carbanak}& A & \%Mail Application\% & Thunderbird \\
       \cline{2-4}
       & B & $*$.\%exe\% & invitation.exe \\
       \cline{2-4}
       & C & $*$ & invitation\\
       \cline{2-4}
       & D & \%system32\%\textbackslash svchost & C:\textbackslash Windows\textbackslash SysWOW64\textbackslash svchost.exe:WofCompressedData\\
       \cline{2-4}
       & E & svchost & svchost\\
       \cline{2-4}
       & F & $*$Sys$\$$ & None\\
       \cline{2-4}
       & G & \%COMMON\_APPDATA\%\textbackslash Mozilla\textbackslash $*$.\%exe\% & C:\textbackslash ProgramData\textbackslash Mozilla\textbackslash BwgWXFhfbVpfWgJfBg.bin\\
       \cline{2-4}
       & H & [HKCU]\textbackslash Software\textbackslash Microsoft\textbackslash Windows\textbackslash CurrentVersion \textbackslash Internet Settings & [HKCU]\textbackslash Software\textbackslash Microsoft\textbackslash Windows\textbackslash CurrentVersion \textbackslash Internet Settings\\
       \cline{2-4}
       & I & \%AppData\%\textbackslash Mozilla\textbackslash Firefox\textbackslash $*$\textbackslash prefs.js & C:\textbackslash Users\textbackslash test\_user\textbackslash AppData\textbackslash Roaming\textbackslash Mozilla\textbackslash Firefox\textbackslash Profiles\textbackslash\- ddl1t72n.default\textbackslash prefs.js\\
       \cline{2-4}
       & J & \%External\nespace IP\nespace address\% & None\\
       \hline
       \multirow{12}{*}{Uroburos}& A & $*$ & contract.exe \\
       \cline{2-4}
       & B & \%APPDATA\%\textbackslash Microsoft\textbackslash credprov.tlb & C:\textbackslash Users\textbackslash test\_user\textbackslash AppData\textbackslash Roaming\textbackslash Microsoft\textbackslash credprov.tlb\\
       \cline{2-4}
       & C & \%APPDATA\%\textbackslash Microsoft\textbackslash shdocvw.tlb & C:\textbackslash Users\textbackslash test\_user\textbackslash AppData\textbackslash Roaming\textbackslash Microsoft\textbackslash shdocvw.tlb \\
       \cline{2-4}
       & D & rundll32 & rundll32\\
       \cline{2-4}
       & E & [HKCU]\textbackslash Software\textbackslash Classes\textbackslash CLSID\textbackslash {42aedc87-2188-41fd-b9a3-0c966feabec1}\textbackslash & [HKCU]\textbackslash Software\textbackslash Classes\textbackslash CLSID\textbackslash {42aedc87-2188-41fd-b9a3-0c966feabec1}\textbackslash\\
       \cline{2-4}
       & F & $*$\textbackslash winview.ocx & None \\
       \cline{2-4}
       & G & $*$\textbackslash mskfp32.ocx & None\\
       \cline{2-4}
       & H & $*$\textbackslash msvcrtd.tlb & None\\
       \cline{2-4}
       & I & \%APPDATA\%\textbackslash Microsoft\textbackslash oleaut32.dll & C:\textbackslash Users\textbackslash test\_user\textbackslash AppData\textbackslash Roaming\textbackslash Microsoft\textbackslash oleaut32.dll\\
       \cline{2-4}
       & J & \%APPDATA\%\textbackslash Microsoft\textbackslash oleaut32.tlb & C:\textbackslash Users\textbackslash test\_user\textbackslash AppData\textbackslash Roaming\textbackslash Microsoft\textbackslash oleaut32.tlb\\
       \cline{2-4}
       & K & \%APPDATA\%\textbackslash Microsoft\textbackslash libadcodec.dll & C:\textbackslash Users\textbackslash test\_user\textbackslash AppData\textbackslash Roaming\textbackslash Microsoft\textbackslash libadcodec.dll\\
       \cline{2-4}
       & L & \%APPDATA\%\textbackslash Microsoft\textbackslash libadcodec.tlb & C:\textbackslash Users\textbackslash test\_user\textbackslash AppData\textbackslash Roaming\textbackslash Microsoft\textbackslash libadcodec.tlb\\
       \hline
       \multirow{9}{*}{DustySky}& A & $*$.\%exe\% & News.docx.exe \\
       \cline{2-4}
       & B & $*$ & News\\
       \cline{2-4}
       & C & \%Microsoft Word\% & C:\textbackslash Program Files\textbackslash Microsoft Office\textbackslash Office12\textbackslash Winword.exe\\
       \cline{2-4}
       & D & $*$\textbackslash vboxmrxnp.dll & C:\textbackslash WINDOWS\textbackslash vboxmrxnp.dlls\\
       \cline{2-4}
       & E & $*$\textbackslash vmbusres.dll & C:\textbackslash WINDOWS\textbackslash vmbusres.dlls\\
       \cline{2-4}
       & F & $*$\textbackslash vmGuestlib.dll & C:\textbackslash WINDOWS\textbackslash SysWOW64\textbackslash vmGuestLib.dll\\
       \cline{2-4}
       & G & \%TEMP\%\textbackslash $*$.\%exe\% & C:\textbackslash Users\textbackslash test\_user\textbackslash AppData\textbackslash Local\textbackslash Temp \textbackslash 1371372533114561232114361100131187183149253.exe\\
       \cline{2-4}
       & H & $*$ & 1371372533114561232114361100131187183149253\\
       \cline{2-4}
       & I & \%TEMP\%\textbackslash temps & C:\textbackslash Users\textbackslash test\_user\textbackslash AppData\textbackslash Local\textbackslash Temp\textbackslash temps \\
       \hline
       
       \multirow{13}{*}{OceanLotus}& A & $*$.\%exe\% & Chi tiet don khieu nai gui saigontel.exe \\
       \cline{2-4}
       & B & $*$ & Chi tiet don khieu nai gui saigontel\\
       \cline{2-4}
       & C & \%temp\%\textbackslash $*$ & C:\textbackslash Users\textbackslash test\_user\textbackslash AppData\textbackslash Local\textbackslash Temp\textbackslash tmp.docx \\
       \cline{2-4}
       & D & \%temp\%\textbackslash [0-9].tmp.\%exe\% & None \\
       \cline{2-4}
       & E & \%Microsoft Word\% & C:\textbackslash Program Files\textbackslash Microsoft Office\textbackslash Office12\textbackslash Winword.exe \\
       \cline{2-4}
       & F & $*$\textbackslash rastlsc.\%exe\% & C:\textbackslash Program Files (x86)\textbackslash Symantec\textbackslash Officewordtask\textbackslash rastlsc.exe\\
       \cline{2-4}
       & G & $*$\textbackslash rastls.dll & C:\textbackslash Program Files (x86)\textbackslash Symantec\textbackslash Officewordtask\textbackslash rastls.dll\\
       \cline{2-4}
       & H & $*$\textbackslash (Sylog.bin\textbar OUTLFLTR.DAT) & C:\textbackslash Program Files (x86)\textbackslash Symantec\textbackslash Officewordtask\textbackslash OUTLFLTR.DAT\\
       \cline{2-4}
       & I & rastlsc & rastlsc\\
       \cline{2-4}
       & J & \textbackslash SOFTWARE\textbackslash Classes\textbackslash AppX$*$ & None\\
       \cline{2-4}
       & K & $*$\textbackslash HTTPProv.dll & None\\
       \cline{2-4}
       & L & SOFTWARE\textbackslash Classes\textbackslash CLSID\textbackslash \{E3517E26-8E93-458D-A6DF-8030BC80528B\} & SOFTWARE\textbackslash Classes\textbackslash CLSID\textbackslash \{E3517E26-8E93-458D-A6DF-8030BC80528B\}\\
       \cline{2-4}
       & M & \%External\nespace IP\nespace address\% & None\\
       \hline
   \end{tabular}
    \normalsize
    \vspace{1em}
   \caption{Node labels of the query graphs in \Cref{fig:etw_4_in_one} and their alignments.}\label{etw-eval-aligns1}
   \vspace{-.7em}
   \end{table*}

 \begin{figure*}
  \begin{center}
    \includegraphics[width=.96\textwidth]{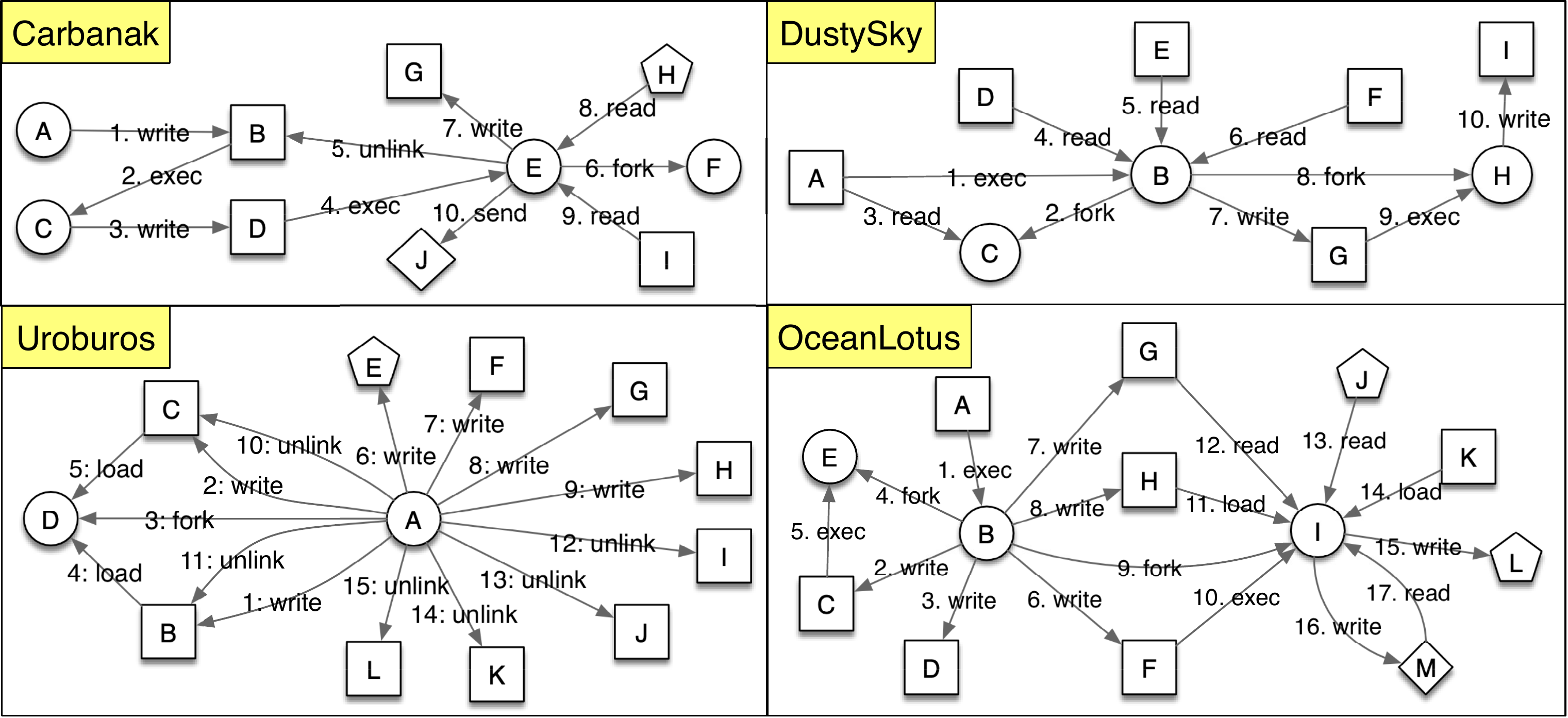}
  \end{center}
  \vspace{.5em}
    \caption{Query Graphs of Carbanak, Uroburos, DustySky, and OceanLotus malware extracted from their CTI reports.
    }
    \label{fig:etw_4_in_one}
\end{figure*}

   \begin{table*}
 \footnotesize
   \begin{tabular}{|M{1.1cm}|M{.5cm}|>{\raggedright\arraybackslash}M{6.3cm}|>{\raggedright\arraybackslash}M{8cm}|}
        \hline
        \textbf{Malware} & \textbf{Node} & \textbf{Label} & \textbf{Aligned Node Label}\\
      \hline  
      \multirow{14}{*}{njRAT}& A & $*$ & Authorization \\
      \cline{2-4}
       & B & $*$.exe.config &C:\textbackslash Users\textbackslash test\_user\textbackslash Desktop\textbackslash Authorization.exe.config \\
       \cline{2-4}
       & C & $*$.tmp & C:\textbackslash Users\textbackslash test\_user\textbackslash AppData\textbackslash Roaming\textbackslash ja33kk.exe.tmp\\
       \cline{2-4}
       & D & C:\textbackslash WINDOWS\textbackslash Prefetch\textbackslash $*$.EXE-$*$.pf & C:\textbackslash Windows\textbackslash Prefetch\textbackslash AUTHORIZATION.EXE-69AD75AA.pf\\
       \cline{2-4}
       & E & \%APPDATA\%\textbackslash $*$ & C:\textbackslash Users\textbackslash test\_user\textbackslash AppData\textbackslash Roaming\textbackslash ja33kk.exe\\
       \cline{2-4}
       & F & $*$ & ja33kk\\
       \cline{2-4}
       & G & C:\textbackslash WINDOWS\textbackslash Prefetch\textbackslash $*$.EXE-$*$.pf & C:\textbackslash Windows\textbackslash Prefetch\textbackslash JA33KK.EXE-7FA5E873.pf\\
       \cline{2-4}
       & H & \%USER\_PROFILE\%\textbackslash Start Menu\textbackslash Programs\textbackslash Startup\textbackslash $*$ & C:\textbackslash Users\textbackslash test\_user\textbackslash AppData\textbackslash Roaming\textbackslash Microsoft\textbackslash Windows\textbackslash Start Menu\textbackslash Programs\textbackslash Startup\textbackslash9758a8dfbe15a00f55a11c8306f80da1.exe\\
       \cline{2-4}
       & I & netsh & netsh \\
       \cline{2-4}
       & J & C:\textbackslash WINDOWS\textbackslash Prefetch\textbackslash NETSH.EXE-$*$.pf & C:\textbackslash Windows\textbackslash Prefetch\textbackslash NETSH.EXE-CD959116.pf\\
       \cline{2-4}
       & K & [HKCU]\textbackslash Software\textbackslash Microsoft\textbackslash Windows\textbackslash CurrentVersion\textbackslash Run\textbackslash & [HKCU]\textbackslash Software\textbackslash Microsoft\textbackslash Windows\textbackslash CurrentVersion\textbackslash Run\textbackslash\\
       \cline{2-4}
       & L & [HKLM]\textbackslash Software\textbackslash Microsoft\textbackslash Windows\textbackslash CurrentVersion\textbackslash Run\textbackslash& [HKLM]\textbackslash Software\textbackslash Microsoft\textbackslash Windows\textbackslash CurrentVersion\textbackslash Run\textbackslash\\
       \cline{2-4}
       & M & [HKLM]\textbackslash SYSTEM\textbackslash CurrentControlSet\textbackslash Services\textbackslash SharedAccess\textbackslash \-Parameters\textbackslash FirewallPolicy\textbackslash
      StandardProfile\textbackslash Authorized\-Applications\textbackslash List\textbackslash APPDATA\textbackslash
        & None \\
       \cline{2-4}
       & N & \%External\nespace IP\nespace address\% & None \\
       \hline
       \multirow{31}{*}{HawkEye}& A & $*$.\%Compressed\% &  PROFORMA INVOICE\nespace  \_20190423072201\nespace  pdf.bin.zip \\
       \cline{2-4}
       & B & \%Unachiever\% & WinRAR.exe\\
       \cline{2-4}
       & C & $*$.\%exe\% & C:\textbackslash Users\textbackslash test\_user\textbackslash Desktop\textbackslash PROFORMA INVOICE\nespace  \_20190423072201\nespace  pdf.bin\\
       \cline{2-4}
       & D & $*$ & PROFORMA INVOICE\nespace  \_20190423072201\nespace  pdf\\
       \cline{2-4}
       & E & RegAsm & RegAsm\\
       \cline{2-4}
       & F & vbc & vbc (PID$_1$)\\
       \cline{2-4}
       & G & vbc & vbc (PID$_2$)\\
       \cline{2-4}
        & H$_1$ & $*$Opera$*$ & C:\textbackslash Users\textbackslash test\_user\textbackslash AppData\textbackslash Roaming\textbackslash Opera\textbackslash Opera7\textbackslash profile\textbackslash wand.dat \\
       \cline{2-4}
       & H$_2$ & $*$Chrome$*$ & C:\textbackslash Users\textbackslash test\_user\textbackslash AppData\textbackslash Local\textbackslash Google\textbackslash Chrome\textbackslash User\nespace Data\textbackslash Default\textbackslash Login\nespace Data\\
       \cline{2-4}
       & H$_3$ & $*$Chromium$*$ &C:\textbackslash Users\textbackslash test\_user\textbackslash AppData\textbackslash Local\textbackslash Chromium\textbackslash User\nespace Data \\
       \cline{2-4}
       & H$_4$ & $*$Chrome SxS$*$ & C:\textbackslash Users\textbackslash test\_user\textbackslash AppData\textbackslash Local\textbackslash Google\textbackslash Chrome\nespace SxS\textbackslash User\nespace Data\\
       \cline{2-4}
       & H$_5$ & $*$Thunderbird$*$ & C:\textbackslash Users\textbackslash test\_user\textbackslash AppData\textbackslash Roaming\textbackslash Thunderbird\textbackslash Profiles\\
       \cline{2-4}
       & H$_6$ & $*$SeaMonkey$*$ & C:\textbackslash Users\textbackslash test\_user\textbackslash AppData\textbackslash Roaming\textbackslash Mozilla\textbackslash SeaMonkey\textbackslash Profiles\\
       \cline{2-4}
       & H$_7$ & $*$SunBird$*$ & None\\
       \cline{2-4}
       & H$_8$ & $*$IE$*$ & C:\textbackslash Users\textbackslash test\_user\textbackslash AppData\textbackslash Local\textbackslash Microsoft\textbackslash Windows\textbackslash History\textbackslash History.IE5\\
       \cline{2-4}
       & H$_9$ & $*$Safari$*$ & None\\
       \cline{2-4}
       & H$_{10}$ & $*$Firefox$*$ & C:\textbackslash Users\textbackslash test\_user\textbackslash AppData\textbackslash Roaming\textbackslash Mozilla\textbackslash Firefox\textbackslash profiles.ini\\
       \cline{2-4}
       & H$_{11}$ & $*$Yandex$*$ & C:\textbackslash Users\textbackslash test\_user\textbackslash AppData\textbackslash Local\textbackslash Yandex\textbackslash YandexBrowser\textbackslash User\nespace Data\textbackslash Default\textbackslash Login\nespace Data\\
       \cline{2-4}
       & H$_{12}$ & $*$Vivaldi$*$ & C:\textbackslash Users\textbackslash test\_user\textbackslash AppData\textbackslash Local\textbackslash Vivaldi\textbackslash User\nespace Data\textbackslash Default\textbackslash Login\nespace Data\\
       \cline{2-4}
       & I$_1$ & $*$Yahoo$*$ & [HKLM]\textbackslash Software\textbackslash Yahoo\textbackslash Pager\\
       \cline{2-4}
       & I$_2$ & $*$GroupMail$*$ & None\\
       \cline{2-4}
       & I$_3$ & $*$Thunderbird$*$ & C:\textbackslash Users\textbackslash test\_user\textbackslash AppData\textbackslash AppData\textbackslash Roaming\textbackslash Thunderbird\textbackslash Profiles\\
       \cline{2-4}
       & I$_4$ & $*$MSNMessenger$*$ & [HKLM]\textbackslash Software\textbackslash Microsoft\textbackslash MSNMessenger\\
       \cline{2-4}
       & I$_5$ & $*$Windows\nespace Mail$*$ & C:\textbackslash Users\textbackslash test\_user\textbackslash AppData\textbackslash Local\textbackslash Microsoft\textbackslash Windows\nespace Mail\\
       \cline{2-4}
       & I$_6$ & $*$IncrediMail$*$ & [HKLM]\textbackslash Software\textbackslash WOW6432Node\textbackslash IncrediMail\textbackslash Identities\\
       \cline{2-4}
       & I$_7$ & $*$Outlook$*$ & [HKLM]\textbackslash Software\textbackslash Microsoft\textbackslash Office\textbackslash 16.0\textbackslash Outlook\textbackslash Profiles\\
       \cline{2-4}
       & I$_8$ & $*$Eudora$*$ & [HKLM]\textbackslash Software\textbackslash Qualcomm\textbackslash Eudora\textbackslash CommandLine\\
       \cline{2-4}
       & J & \%temp\%\textbackslash $*$.tmp & C:\textbackslash Users\textbackslash test\_user\textbackslash AppData\textbackslash Local\textbackslash Temp\textbackslash tmp8FC3.tmp\\
       \cline{2-4}
       & K & \%temp\%\textbackslash $*$.tmp & C:\textbackslash Users\textbackslash test\_user\textbackslash AppData\textbackslash Local\textbackslash Temp\textbackslash tmp8BAB.tmp\\
       \cline{2-4}
       & L & http[s]:\textbackslash\textbackslash whatismyipaddress.com\textbackslash $*$ & None \\
       \cline{2-4}
        & M & \%External\nespace IP\nespace address\% & None\\
       \hline
        \multirow{4}{*}{DeputyDog}& A & $*$.\%exe\% & C:\textbackslash Users\textbackslash test\_user\textbackslash Desktop\textbackslash img20130823.jpg.exe \\
       \cline{2-4}
       \multirow{4}{*}{(\Cref{fig:deputydog})}& B & $*$ & img20130823\\
       \cline{2-4}
       & C & \%APPDATA\%\textbackslash $*$ & C:\textbackslash ProgramData\textbackslash 28542CC0.dll\\
       \cline{2-4}
       & D & [HKCU]\textbackslash Software\textbackslash Microsoft\textbackslash Windows\textbackslash CurrentVersion\textbackslash\- Run\textbackslash & None\\
       \cline{2-4}
       & E & \%External\nespace IP\nespace address\% & 180.150.228.102\\
       \hline
   \end{tabular}
   \normalsize
    \vspace{1em}
 \caption{Node labels of the query graphs in  \Cref{fig:deputydog,fig:etw_2_in_one}  and their alignments.}\label{etw-eval-aligns2}
\end{table*}

 \begin{figure*}
  \begin{center}
    \includegraphics[width=.96\textwidth]{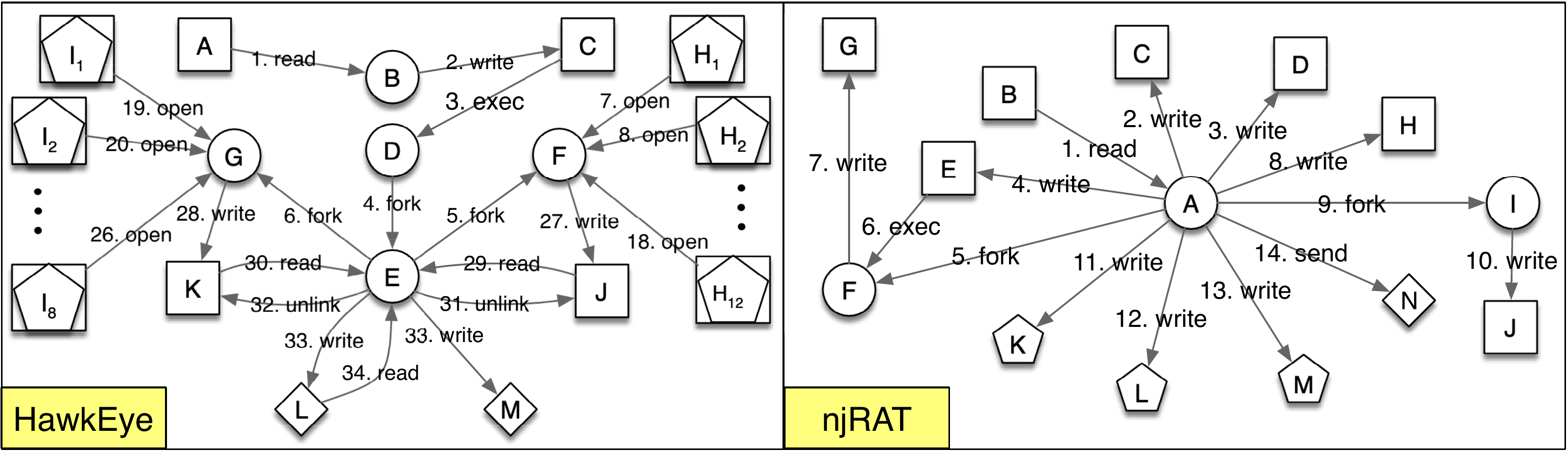}
  \end{center}
  \vspace{1em}
    \caption{Query Graphs of HawkEye and njRAT malware, extracted from their CTI reports.
    }
    \label{fig:etw_2_in_one}
\end{figure*} 
\endgroup
} %

\noindent
\textbf{Mutation Detection Evaluation.}
As mentioned earlier, a common practice among attackers is that of mutating malware to evade detection or to add more features to it. 
Therefore, a CTI report may describe the behavior of a different version of the malware that is actually present in the system, and it is vital for a threat hunting tool to be able to detect different mutations of a malware sample.
To this end, we execute several real-world malware families, containing different mutated versions of the same malware, in a controlled environment.  
The fourth column of \cref{malware-db}, shows the number of  malware samples with different hash values belonging to the family mentioned in the corresponding CTI report. We note that the reports describe the behavior of only a few samples.
The fifth column of \cref{malware-db} shows our selected sample's hash value, while the sixth column shows the relation between our selected sample and the ones the CTI report is based on.
For instance, the reports of DeputyDog, Uroburos, and OceanLotus  cover different activities performed by a set of different samples, and our selected sample is one of them. We have aggregated all those activities in one query graph.
For the other test cases, the sample we have executed is different from the ones that the report is based on, which could be considered as detecting a mutated malware.
njRAT and DustySky  explicitly mention their analyzed sample, which are different from the one we have chosen.
The Carbanak report mentions 109 samples, from which we have randomly selected one.
Finally, the sample of HawkEye malware is selected from an external source and is not among the samples mentioned in the report.

\noindent
\textbf{Comparison with Existing Tools.}
We compare \projname with the results of three other tools, namely RedLine\cite{redline}, Loki\cite{loki}, and Splunk\cite{splunk}.
The input to these tools is extracted from the same report we extract the query graphs and contains IOCs in different types such as hash values, process names, file names, and registries. 
We have transformed these IOCs to the accepted format of each tool (e.g., RedLine accepts input in OpenIOC format \cite{open-ioc}).
The number of IOCs submitted to Redline, Loki, and Splunk are shown in column-7, while the query graphs submitted to \projname are shown in 
\Cref{fig:etw_4_in_one,fig:etw_2_in_one}. 
A detailed explanation of these query graphs demonstrating how they are constructed can be found in \cref{malware_description}.
The correspondence between node labels in the query graphs and their actual names is represented in the second and third columns of \cref{etw-eval-aligns1,etw-eval-aligns2}, while the alignments produced by \projname are shown in the
last column.

As shown in 
the extracted query graphs,
the design of \projname~'s search method, which is based on the information flow and causal dependencies, makes us capable to include benign nodes (nodes C, D, E, and F in DustySky) or attack nodes with exact same names of benign entities (node E in Carbanak) in the query graph.
However, these entity names could not be defined as an IOC in the other tested tools as will lead to many false positive alarms.
As Redline, Loki, and Splunk look for each IOC in isolation, they expect IOCs as input that are always malicious regardless of their connection with other entities.
To this end, we do a preliminary search for each isolated  IOC in a clean system and make sure that we have only extracted IOCs that have no match in a clean system.
As a result, for some test cases, like HawkEye, although the behavior graph is rich,
there are not so many isolated IOCs except a few hash values that could be defined.
This highlights the importance of the dependencies between IOCs, which is the foundation of \projname's search algorithm, and is not considered by other
tools.

\noindent
\textbf{Detection Results.}
The last four columns of \cref{malware-db} contain the detection results, which show how each tool could detect the tested samples.
Keywords B, F, H, P, and R represent detection based on the behavior, file name, hash value, process name, and registry, respectively.
In addition, some of the tested tools feature other methods to detect anomalies, injection, or other security incidents.
Among these, we encountered some alarms from
Windows Security Mitigation and
PE-Sieve \cite{pe-sieve}, which are represented by keywords S and PE, respectively.
While for \projname, a score is shown which shows the goodness of the overall alignment of each query graph, for the other tools, $\times N$ indicates the number of hits when there has been more than one hit for a specific type of IOC.

As shown in \cref{malware-db}, for all the test cases,
\projname has found an alignment that bypasses the threshold value of $\frac{1}{3}$ .
After running the search algorithm, in most of the cases, \projname found a node alignment for only a subset of the entities in the query graph,
except for DustySky, where \projname  found a node alignment for every  entity.
The information flows and causal dependencies that appear among the aligned nodes are often the same as the query graph with some exceptions. For example, in contrast to how it appears in the query graph of njRAT, where node A generates most of the actions, in our experiment, node F generated most of the actions, such as  the write event to nodes C, G, K, L, and the fork event of node I.
However, since there is a path from node A to node F in the query graph, \projname was  able to find these alignments and measure a high alignment score.

The samples of njRAT, DeputyDog, Uroburous, and OceanLotus are also detected by all the other tools, as these samples use unique names or hash values that are available in the threat intelligence and could be attributed to those malwares.
For the other three test cases, none of the isolated IOCs could be detected because of different reasons such as malware mutations, using random names in each run (nodes J and K in HawkEye query graph), and using legitimate libraries or their similar names.
Nevertheless, Splunk found an ETW event related to the Carbanak sample, which is generated when Windows Security Mitigation service has blocked svchost from generating dynamic code.
Loki's PE-Sieve has also detected some attempts of code implants which have resulted in raising some warning signal and not an alert. 
PE-Sieve detects all modifications done to a process even though they may not necessarily be malicious.
As such modifications happen regularly in many benign scenarios, PE-Sieve detections are considered as warning signals that need further investigations.

\noindent
\textbf{Conclusions.}
Our analysis results show that other tools usually perform well when the sample is a subset of the ones the report is written based on. 
This situation is similar to when there is no mutations, and therefore, there are unique hash values or names that could be used as signature of a malware.
For example, DeputyDog sample drops many files with unique names and hash values that do not appear in a benign system, and finding them is a strong indication of this malware.
However, its query graph (\Cref{fig:deputydog}) is not very rich, and \projname has not been able to correlate the modified registry (node D) with the rest of the aligned nodes. 
Although the calculated score is still higher than the  threshold, but the other tools might perform better when the malware is using well-known IOCs that are
strong enough to indicate an attack in isolation.

On the contrary, when the chosen sample is different from the samples analyzed by the report, which is similar to the case that malware is mutated, other tools usually are not able to find the attacks.
In such situations, \projname has a better chance to detect the attack as the behavior often remains constant among the mutations.

\begin{table*}[!ht]
 \renewcommand\thetable{8}
 \footnotesize
 \begin{center}

   \begin{tabular}{|M{1.1cm}|M{1.8cm}|M{1.5cm}|M{1.2cm}|M{1.7cm}|M{1.1cm}|M{1.1cm}|M{1cm}|>{\raggedright\arraybackslash}M{3.5cm}|}
      \hline
      \textbf{Scenario} & \textbf{Size on Disk (Uncompressed)} & \textbf{Consumption time} & \textbf{Occupied Memory} & \textbf{Log Duration} & \textbf{sub $\in$ \textbar $V(G_p)$\textbar} & \textbf{obj $\in$ \textbar $V(G_p)$\textbar} & \textbf{\textbar $E(G_p)$\textbar} & \textbf{Search Time (s)}\\
      \hline
      BSD-1 & 3022 MB & 0h-34m-59s  & 867 MB & 03d-18h-01m & 110.66 K & 1.48 M & 7.53 M & 3.28\\
      \hline
      BSD-2 & 4808 MB &  0h-58m-05s  & 1240 MB & 05d-01h-15m  & 213.10 K & 2.25 M & 12.66 M &0.04\\
      \hline
      BSD-3\&4 & 1828 MB &  0h-21m-31s  & 638 MB & 02d-00h-59m  & 84.39 K & 897.63 K & 4.65 M & 26.09 (BSD-3), 1.47 (BSD-4)\\
      \hline
      Win-1\&2 & 54.57 GB & 4h-58m-30s  & 3790 MB & 08d-13h-35m & 1.04 M & 2.38 M & 70.82 M & 125.26 (Win-1), 46.02 (Win-2)\\
      \hline  
      Linux-1\&2 & 9436 MB & 1h-26m-37s  & 4444 MB & 03d-04h-20m  & 324.68 K & 30.33 M & 51.98 M & 1279.32 (Linux-1), 1170.86 (Linux-2)\\
      \hline  
      Linux-3 & 131.1 GB & 2h-30m-37s  & 21.2 GB & 10d-15h-52m  & 374.71 K & 5.32 M & 69.89 M & 385.16\\
      \hline
      Linux-4 & 4952 MB & 0h-04m-00s  & 1095 MB & 00d-07h-13m & 35.81 K & 859.03 K & 13.06 M & 20.72\\
      \hline

   \end{tabular}
 \end{center}
 \normalsize
\vspace{1em}
 \caption{Statistics of logs, Consumption and Search Times.}\label{tab:consumptions}
\end{table*}

\subsection{Evaluation on Benign Datasets}

To stress-test \projname on false positives, we used the benign dataset generated as part of the adversarial engagement in the DARPA TC program and four machines (a client, a SSH server, a mail server and a web server) we monitored for one month.
Collectively, these datasets contained over seven months worth of benign audit records and billions of audit records on Windows, Linux, and FreeBSD.
During this time, multiple users used these systems and typical attack-free actions were conducted including web browsing, installing security updates (including kernel updates), virus scanning, taking backups, and software uninstalls.

After collecting the logs, we run \projname to construct the provenance graph, and then search for all the query graphs we have extracted from the TC reports and the public malware reports.
We try up to 20 iterations starting from different seed node selections per each query graph per each provenance graph and select the highest score. 
Note that although these logs are attack-free, they share many nodes and events with our query graphs, such as confidential files, critical system files, file editing tools, or processes related to web browsing/hosting, and email clients, all of which were accessed during the benign data collection period.
However, even in cases where similar flows appear by chance, the {\em influence score} prunes away many of these flows. Consequently, the graph alignment score \projname calculates among all the benign datasets is at most equal to 0.16, well below the threshold.

\noindent
\textbf{Validating the Threshold Value.}\label{validating_threshold}
\begin{figure}[t]
  \begin{center}

    \includegraphics[width=\columnwidth]{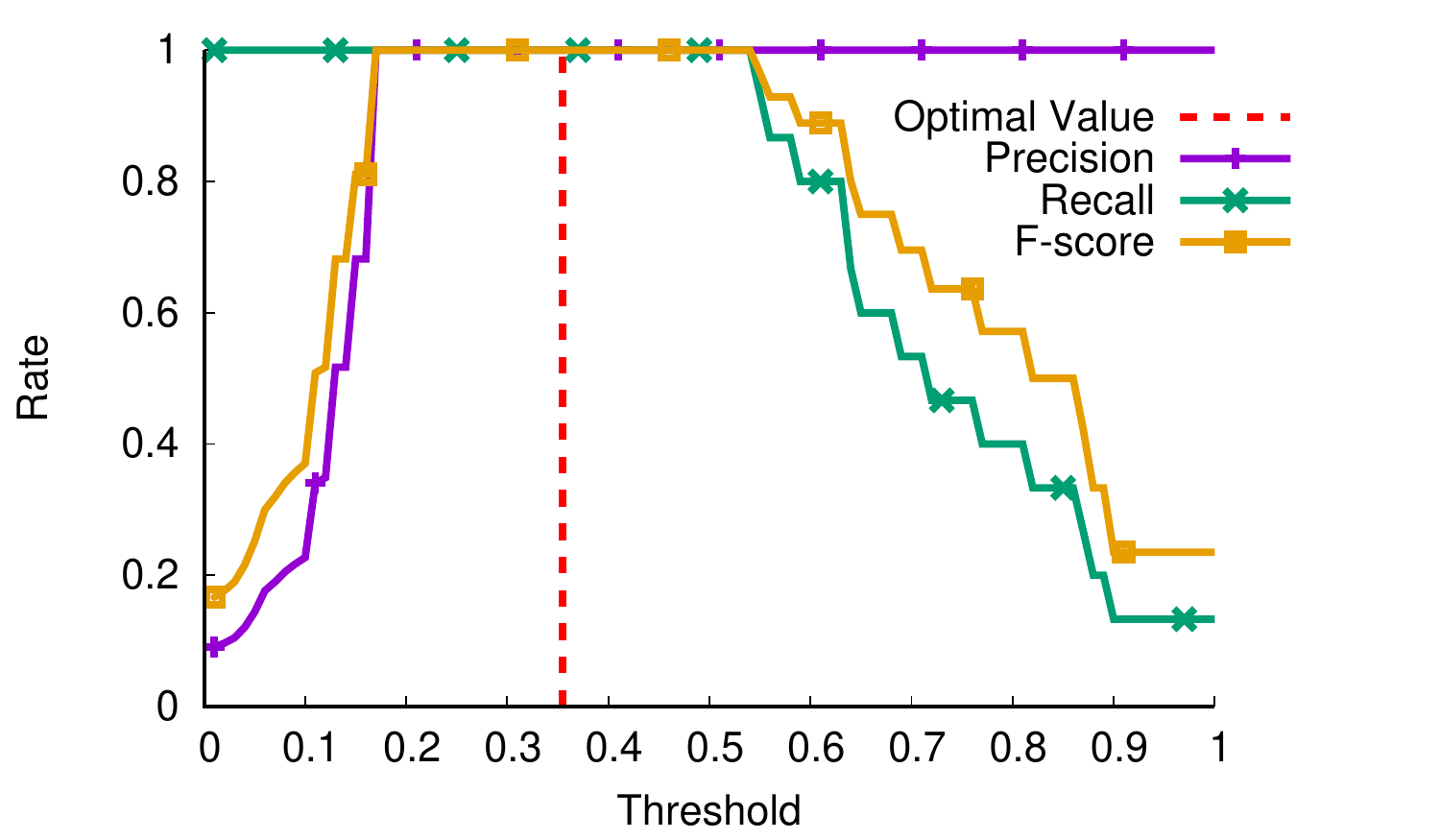}
  \end{center}
    \caption{Selecting the Optimal Threshold Value.}
    \label{fig:fscore}
\end{figure} 
The selection of the threshold value is critical to avoid false signals.
Too low a threshold could result in premature matching (false positives) while too high a threshold could lead to missing reasonable matches (false negatives).
Thus, there is a trade-off in choosing an optimal threshold value.
To determine the optimal threshold value, we measured the {\em F-score} using varying threshold values, as shown in \Cref{fig:fscore}.
This analysis is done based on the highest alignment score calculated in 20 iterations of \projname's search algorithm for all the attack and benign scenarios we have evaluated.
As it is shown, the highest F-score value is achieved when the threshold is at the interval [0.17, 0.54], which is the range in which all attack subgraphs are correctly found, and no alarm is raised for benign datasets. 
The middle of this interval, i.e., 0.35, maximizes the margin between attack and benign scores, and choosing this value as the optimal threshold minimizes the classification errors,.
Therefore, we set the $C_{thr}$ to $3$ which results in $\frac{1}{C_{thr}} = \frac{1}{3}$ which is 
close to the optimal value.

\subsection{Efficiency}\label{eval_effic}

\begin{table}[b]
\renewcommand\thetable{7}
 \footnotesize
 \begin{center}
    \begin{tabular}{|M{1.1cm}|M{0.7cm}|M{1cm}|M{1cm}|M{1cm}|M{1.4cm}|}
      \hline
      \textbf{Detection} & \multirow{ 2}{*}{\textbf{Type}} & \multicolumn{3}{|c|}{\textbf{Runtime Overhead}} & \textbf{Search} \\
      \cline  {3-5}
      \textbf{Method} & & \textbf{Apache \cite{apache-b}} & \textbf{JetStream \cite{jetstream}} & \textbf{HDTune \cite{hdtune}} & \textbf{Time (min)} \\
      \hline
      Redline & offline & - & - & - & 124 \\
      \hline    
      Loki & offline & - & - & - & 215\\
      \hline
      Splunk & online & 3.70\% & 2.94\% & 4.37\% & < 1\\
      \hline
      \projname & online & 0.82\% & 1.86\% & 0.64\% & < 1\\
      \hline
    \end{tabular}
 \end{center}
  \normalsize
  \vspace*{1em}
  \caption{Efficiency Comparison with Related Systems.}
  \label{tab:perf}
\end{table}

The overheads and search times for the different tools we used are shown in \cref{tab:perf}.
Redline and Loki are offline tools, searching for artifacts that are left by the attacker on the system, while Splunk and \projname are online tools, searching based on system events collected during runtime.
Hence, Redline and Loki have no runtime overhead due to audit log collection.
The runtime overheads of Splunk and \projname due to log collection are measured using Apache benchmark \cite{apache-b}, which measures web server responsiveness, JetStream \cite{jetstream}, which measures browser execution times, and HDTune \cite{hdtune}, which measures heavy hard drive transactions.
As shown in \cref{tab:perf}, both tools have shown negligible runtime overhead, while the  runtime of Splunk can  be further improved by setting it up in a distributed setting and offloading the data indexing task to another server.

The last column of \cref{tab:perf} shows the time it took searching for IOCs per each tool.
The search time of offline tools highly depends on the number of running processes and volume of occupied disk space, which was 500 GB in our case.
On the other hand, the search time of online methods highly depends on the log size, type and number of activities represented by the logs.
As our experiments with real-world malware samples were running in a controlled environment without many background benign activities and  Internet connection, 
both Splunk and \projname spend less than one minute to search for all the IOCs mentioned in \cref{malware-db}.
In the following, we perform an in-depth analysis of \projname~'s efficiency on the DARPA TC scenarios, which overall contain over a month worth of log data with combined attack and benign activities.
The analysis is done on an 8-core CPU with a 2.5GHz speed each and a 150GB of RAM.

\noindent
\textbf{Audit Logs Consumption.}
In \cref{tab:consumptions}, the second column shows the initial size of the logs on disk, the third column represents
the time it takes to consume all audits log events from disk for building the provenance graph in memory.
This time is measured as the wall-clock time and varies depending on the size of each audit log and the structure of audits logs generated in each platform (BSD, Windows, Linux).
The fourth column shows the total memory consumption by each provenance graph. 
Comparing the size on disk versus memory, we notice that we have an average compression of 1:4 (25\%) via a compact in-memory provenance graph representation based on\cite{hossain2017sleuth}.
However, if memory is a concern, it is still possible to achieve better compression using additional techniques proposed in this area \cite{lee2013loggc,xu2016high,depPresRed18}. The fifth column shows the duration during which the logs were collected while columns 6, 7, and 8 show the total number of subjects (i.e. processes), objects, and events in the provenance graph that is built from the logs, respectively. 
We note that the {\em query graphs} are on average 209K times smaller than the provenance graph for these scenarios. 
Nevertheless,  \projname is still able to find the exact embedding of $G_q$ in $G_p$ very fast, as shown in the last column. We note that some scenarios are joined (e.g., Win-1\&2) because they were executed concurrently on the same machines. 

\noindent
\textbf{Graph Analytics.}
In the last column of \cref{tab:consumptions}, we show the runtime of graph analytics for \projname's search algorithm.
These times are measured from the moment a search query is submitted until we find a similar graph in $G_p$ with an alignment score that surpasses the threshold.
Therefore, for Linux-2, the time includes the sum of the times for two iterations.
The main bottleneck is on the graph search expansion (Step 3), and the time \projname spends on graph search depends on several factors. Obviously, the sizes of both query and provenance graph are proportional to the runtime. However, we notice that the node names in $G_q$ and the shape of this graph have a more significant effect. In particular, when there are nodes with many candidate alignments, there is a higher chance to reverse the direction multiple times and runtime increases accordingly. 

\section{Conclusion}\label{sec:conclusion}
\projname formulates cyber threat hunting as a graph pattern matching problem to
reliably detect known cyber attacks. 
\projname is based on an efficient alignment algorithm to find an embedding of a graph representing the threat behavior in the provenance graph of kernel audit records. We evaluate \projname on real-world cyber attacks and on ten attack scenarios conducted by a professional red-team, over three OS platforms, with tens of millions of audit records. \projname  successfully detects all the attacks with high confidence, no false signals, and in a matter of minutes. 

\begin{acks}
This work was supported by DARPA under  SPAWAR (N6600118C4035), AFOSR (FA8650-15-C-7561), and NSF (CNS-1514472, CNS-1918542 and DGE-1069311). The views, opinions, and/or findings expressed are those of the authors and should not be interpreted as representing the official views or policies of the U.S. Government.
\end{acks}

\FloatBarrier
\vspace{1em}
\bibliographystyle{ACM-Reference-Format}
\bibliography{references}


\begin{thebibliography}{78}

%%% the Web address.

\ifx \showCODEN    \undefined \def \showCODEN     #1{\unskip}     \fi
\ifx \showDOI      \undefined \def \showDOI       #1{#1}\fi
\ifx \showISBNx    \undefined \def \showISBNx     #1{\unskip}     \fi
\ifx \showISBNxiii \undefined \def \showISBNxiii  #1{\unskip}     \fi
\ifx \showISSN     \undefined \def \showISSN      #1{\unskip}     \fi
\ifx \showLCCN     \undefined \def \showLCCN      #1{\unskip}     \fi
\ifx \shownote     \undefined \def \shownote      #1{#1}          \fi
\ifx \showarticletitle \undefined \def \showarticletitle #1{#1}   \fi
\ifx \showURL      \undefined \def \showURL       {\relax}        \fi
% invisible to TeX
\providecommand\bibfield[2]{#2}
\providecommand\bibinfo[2]{#2}
\providecommand\natexlab[1]{#1}
\providecommand\showeprint[2][]{arXiv:#2}

\bibitem[\protect\citeauthoryear{Antonakakis, Perdisci, Lee, Vasiloglou, and
  Dagon}{Antonakakis et~al\mbox{.}}{2011}]%
        {antonakakis2011detecting}
\bibfield{author}{\bibinfo{person}{Manos Antonakakis}, \bibinfo{person}{Roberto
  Perdisci}, \bibinfo{person}{Wenke Lee}, \bibinfo{person}{Nikolaos
  Vasiloglou}, {and} \bibinfo{person}{David Dagon}.}
  \bibinfo{year}{2011}\natexlab{}.
\newblock \showarticletitle{Detecting Malware Domains at the Upper DNS
  Hierarchy.}. In \bibinfo{booktitle}{\emph{USENIX security symposium}},
  Vol.~\bibinfo{volume}{11}. \bibinfo{pages}{1--16}.
\newblock


\bibitem[\protect\citeauthoryear{Antonakakis, Perdisci, Nadji, Vasiloglou,
  Abu-Nimeh, Lee, and Dagon}{Antonakakis et~al\mbox{.}}{2012}]%
        {antonakakis2012throw}
\bibfield{author}{\bibinfo{person}{Manos Antonakakis}, \bibinfo{person}{Roberto
  Perdisci}, \bibinfo{person}{Yacin Nadji}, \bibinfo{person}{Nikolaos
  Vasiloglou}, \bibinfo{person}{Saeed Abu-Nimeh}, \bibinfo{person}{Wenke Lee},
  {and} \bibinfo{person}{David Dagon}.} \bibinfo{year}{2012}\natexlab{}.
\newblock \showarticletitle{From Throw-Away Traffic to Bots: Detecting the Rise
  of DGA-Based Malware.}. In \bibinfo{booktitle}{\emph{USENIX security
  symposium}}, Vol.~\bibinfo{volume}{12}.
\newblock


\bibitem[\protect\citeauthoryear{Apache}{Apache}{2019}]%
        {apache-b}
\bibfield{author}{\bibinfo{person}{Apache}.} \bibinfo{year}{2019}\natexlab{}.
\newblock \bibinfo{title}{{ab - Apache HTTP server benchmarking tool}}.
\newblock
  \bibinfo{howpublished}{\url{https://httpd.apache.org/docs/2.4/programs/ab.html}}.
\newblock
\newblock
\shownote{Accessed: 2019-08-27.}


\bibitem[\protect\citeauthoryear{Bilge, Balzarotti, Robertson, Kirda, and
  Kruegel}{Bilge et~al\mbox{.}}{2012}]%
        {bilge2012disclosure}
\bibfield{author}{\bibinfo{person}{Leyla Bilge}, \bibinfo{person}{Davide
  Balzarotti}, \bibinfo{person}{William Robertson}, \bibinfo{person}{Engin
  Kirda}, {and} \bibinfo{person}{Christopher Kruegel}.}
  \bibinfo{year}{2012}\natexlab{}.
\newblock \showarticletitle{Disclosure: detecting botnet command and control
  servers through large-scale netflow analysis}. In
  \bibinfo{booktitle}{\emph{Proceedings of the 28th Annual Computer Security
  Applications Conference}}. ACM, \bibinfo{pages}{129--138}.
\newblock


\bibitem[\protect\citeauthoryear{Blog}{Blog}{2013}]%
        {uroburos-report}
\bibfield{author}{\bibinfo{person}{G~Data Blog}.}
  \bibinfo{year}{2013}\natexlab{}.
\newblock \bibinfo{title}{{The Uroburos case: new sophisticated RAT
  identified}}.
\newblock
  \bibinfo{howpublished}{\url{https://www.gdatasoftware.com/blog/2014/11/23937-the-uroburos-case-new-sophisticated-rat-identified}}.
\newblock
\newblock
\shownote{Accessed: 2019-04-19.}


\bibitem[\protect\citeauthoryear{by~ESET}{by~ESET}{2018}]%
        {oceanlotus-report}
\bibfield{author}{\bibinfo{person}{WeLiveSecurity by ESET}.}
  \bibinfo{year}{2018}\natexlab{}.
\newblock \bibinfo{title}{{OceanLotus: Old techniques, new backdoor}}.
\newblock
  \bibinfo{howpublished}{\url{https://www.welivesecurity.com/wp-content/uploads/2018/03/ESET_OceanLotus.pdf}}.
\newblock
\newblock
\shownote{Accessed: 2019-08-12.}


\bibitem[\protect\citeauthoryear{by~FortiGuard~Labs}{by~FortiGuard~Labs}{2019}]%
        {hawkeye-report}
\bibfield{author}{\bibinfo{person}{Threat~Analysis by FortiGuard~Labs}.}
  \bibinfo{year}{2019}\natexlab{}.
\newblock \bibinfo{title}{{Analysis of a New HawkEye Variant}}.
\newblock
  \bibinfo{howpublished}{\url{https://www.fortinet.com/blog/threat-research/hawkeye-malware-analysis.html}}.
\newblock
\newblock
\shownote{Accessed: 2019-08-12.}


\bibitem[\protect\citeauthoryear{Cheng, Yu, Ding, Philip, and Wang}{Cheng
  et~al\mbox{.}}{2008}]%
        {cheng2008fast}
\bibfield{author}{\bibinfo{person}{Jiefeng Cheng}, \bibinfo{person}{Jeffrey~Xu
  Yu}, \bibinfo{person}{Bolin Ding}, \bibinfo{person}{S~Yu Philip}, {and}
  \bibinfo{person}{Haixun Wang}.} \bibinfo{year}{2008}\natexlab{}.
\newblock \showarticletitle{Fast graph pattern matching}. In
  \bibinfo{booktitle}{\emph{2008 IEEE 24th International Conference on Data
  Engineering}}. IEEE, \bibinfo{pages}{913--922}.
\newblock


\bibitem[\protect\citeauthoryear{Christodorescu, Jha, and
  Kruegel}{Christodorescu et~al\mbox{.}}{2007}]%
        {christodorescu2007mining}
\bibfield{author}{\bibinfo{person}{Mihai Christodorescu},
  \bibinfo{person}{Somesh Jha}, {and} \bibinfo{person}{Christopher Kruegel}.}
  \bibinfo{year}{2007}\natexlab{}.
\newblock \showarticletitle{Mining specifications of malicious behavior}. In
  \bibinfo{booktitle}{\emph{Proceedings of the the 6th joint meeting of the
  European software engineering conference and the ACM SIGSOFT symposium on The
  foundations of software engineering}}. ACM, \bibinfo{pages}{5--14}.
\newblock


\bibitem[\protect\citeauthoryear{De~Nardo, Ranzato, and Tapparo}{De~Nardo
  et~al\mbox{.}}{2009}]%
        {de2009subgraph}
\bibfield{author}{\bibinfo{person}{Lorenzo De~Nardo},
  \bibinfo{person}{Francesco Ranzato}, {and} \bibinfo{person}{Francesco
  Tapparo}.} \bibinfo{year}{2009}\natexlab{}.
\newblock \showarticletitle{The subgraph similarity problem}.
\newblock \bibinfo{journal}{\emph{IEEE Transactions on Knowledge and Data
  Engineering}} \bibinfo{volume}{21}, \bibinfo{number}{5}
  (\bibinfo{year}{2009}), \bibinfo{pages}{748--749}.
\newblock


\bibitem[\protect\citeauthoryear{EFD}{EFD}{2019}]%
        {hdtune}
\bibfield{author}{\bibinfo{person}{EFD}.} \bibinfo{year}{2019}\natexlab{}.
\newblock \bibinfo{title}{{HD Tune}}.
\newblock \bibinfo{howpublished}{\url{https://www.hdtune.com}}.
\newblock
\newblock
\shownote{Accessed: 2019-08-27.}


\bibitem[\protect\citeauthoryear{Fan, Li, Ma, Tang, Wu, and Wu}{Fan
  et~al\mbox{.}}{2010}]%
        {fan2010graph}
\bibfield{author}{\bibinfo{person}{Wenfei Fan}, \bibinfo{person}{Jianzhong Li},
  \bibinfo{person}{Shuai Ma}, \bibinfo{person}{Nan Tang},
  \bibinfo{person}{Yinghui Wu}, {and} \bibinfo{person}{Yunpeng Wu}.}
  \bibinfo{year}{2010}\natexlab{}.
\newblock \showarticletitle{Graph pattern matching: from intractable to
  polynomial time}.
\newblock \bibinfo{journal}{\emph{Proceedings of the VLDB Endowment}}
  \bibinfo{volume}{3}, \bibinfo{number}{1-2} (\bibinfo{year}{2010}),
  \bibinfo{pages}{264--275}.
\newblock


\bibitem[\protect\citeauthoryear{FireEye}{FireEye}{2013}]%
        {open-ioc-use-case}
\bibfield{author}{\bibinfo{person}{FireEye}.} \bibinfo{year}{2013}\natexlab{}.
\newblock \bibinfo{title}{{OpenIOC Series: Investigating with Indicators of
  Compromise (IOCs) - Part I}}.
\newblock
  \bibinfo{howpublished}{\url{https://www.fireeye.com/blog/threat-research/2013/12/openioc-series-investigating-indicators-compromise-iocs.html}}.
\newblock


\bibitem[\protect\citeauthoryear{FireEye}{FireEye}{2018a}]%
        {open-ioc}
\bibfield{author}{\bibinfo{person}{FireEye}.} \bibinfo{year}{2018}\natexlab{a}.
\newblock \bibinfo{title}{{Open IOC}}.
\newblock \bibinfo{howpublished}{\url{https://openioc.org}}.
\newblock


\bibitem[\protect\citeauthoryear{FireEye}{FireEye}{2018b}]%
        {redline}
\bibfield{author}{\bibinfo{person}{FireEye}.} \bibinfo{year}{2018}\natexlab{b}.
\newblock \bibinfo{title}{{Redline}}.
\newblock
  \bibinfo{howpublished}{\url{https://www.fireeye.com/services/freeware/redline.html}}.
\newblock
\newblock
\shownote{Accessed: 2019-04-23.}


\bibitem[\protect\citeauthoryear{Gallagher}{Gallagher}{2006}]%
        {gallagher2006matching}
\bibfield{author}{\bibinfo{person}{Brian Gallagher}.}
  \bibinfo{year}{2006}\natexlab{}.
\newblock \showarticletitle{Matching structure and semantics: A survey on
  graph-based pattern matching}.
\newblock \bibinfo{journal}{\emph{AAAI FS}}  \bibinfo{volume}{6}
  (\bibinfo{year}{2006}), \bibinfo{pages}{45--53}.
\newblock


\bibitem[\protect\citeauthoryear{Gao, Xiao, Li, Li, Jee, Wu, Kim, Kulkarni, and
  Mittal}{Gao et~al\mbox{.}}{2018a}]%
        {gao2018saql}
\bibfield{author}{\bibinfo{person}{Peng Gao}, \bibinfo{person}{Xusheng Xiao},
  \bibinfo{person}{Ding Li}, \bibinfo{person}{Zhichun Li},
  \bibinfo{person}{Kangkook Jee}, \bibinfo{person}{Zhenyu Wu},
  \bibinfo{person}{Chung~Hwan Kim}, \bibinfo{person}{Sanjeev~R Kulkarni}, {and}
  \bibinfo{person}{Prateek Mittal}.} \bibinfo{year}{2018}\natexlab{a}.
\newblock \showarticletitle{$\{$SAQL$\}$: A Stream-based Query System for
  Real-Time Abnormal System Behavior Detection}. In
  \bibinfo{booktitle}{\emph{27th $\{$USENIX$\}$ Security Symposium
  ($\{$USENIX$\}$ Security 18)}}. \bibinfo{pages}{639--656}.
\newblock


\bibitem[\protect\citeauthoryear{Gao, Xiao, Li, Xu, Kulkarni, and Mittal}{Gao
  et~al\mbox{.}}{2018b}]%
        {gao2018aiql}
\bibfield{author}{\bibinfo{person}{Peng Gao}, \bibinfo{person}{Xusheng Xiao},
  \bibinfo{person}{Zhichun Li}, \bibinfo{person}{Fengyuan Xu},
  \bibinfo{person}{Sanjeev~R Kulkarni}, {and} \bibinfo{person}{Prateek
  Mittal}.} \bibinfo{year}{2018}\natexlab{b}.
\newblock \showarticletitle{$\{$AIQL$\}$: Enabling Efficient Attack
  Investigation from System Monitoring Data}. In \bibinfo{booktitle}{\emph{2018
  $\{$USENIX$\}$ Annual Technical Conference ($\{$USENIX$\} \{$ATC$\}$ 18)}}.
  \bibinfo{pages}{113--126}.
\newblock


\bibitem[\protect\citeauthoryear{Giugno and Shasha}{Giugno and Shasha}{2002}]%
        {giugno2002graphgrep}
\bibfield{author}{\bibinfo{person}{Rosalba Giugno} {and}
  \bibinfo{person}{Dennis Shasha}.} \bibinfo{year}{2002}\natexlab{}.
\newblock \showarticletitle{Graphgrep: A fast and universal method for querying
  graphs}. In \bibinfo{booktitle}{\emph{Pattern Recognition, 2002. Proceedings.
  16th International Conference on}}, Vol.~\bibinfo{volume}{2}. IEEE,
  \bibinfo{pages}{112--115}.
\newblock


\bibitem[\protect\citeauthoryear{Goel, Feng, Maier, Feng, and Walpole}{Goel
  et~al\mbox{.}}{2005a}]%
        {forensix}
\bibfield{author}{\bibinfo{person}{A. Goel}, \bibinfo{person}{W.~C. Feng},
  \bibinfo{person}{D. Maier}, \bibinfo{person}{W.~C. Feng}, {and}
  \bibinfo{person}{J. Walpole}.} \bibinfo{year}{2005}\natexlab{a}.
\newblock \showarticletitle{Forensix: a robust, high-performance reconstruction
  system}. In \bibinfo{booktitle}{\emph{25th IEEE International Conference on
  Distributed Computing Systems Workshops}}.
\newblock


\bibitem[\protect\citeauthoryear{Goel, Po, Farhadi, Li, and de~Lara}{Goel
  et~al\mbox{.}}{2005b}]%
        {taser2005}
\bibfield{author}{\bibinfo{person}{Ashvin Goel}, \bibinfo{person}{Kenneth Po},
  \bibinfo{person}{Kamran Farhadi}, \bibinfo{person}{Zheng Li}, {and}
  \bibinfo{person}{Eyal de Lara}.} \bibinfo{year}{2005}\natexlab{b}.
\newblock \showarticletitle{The Taser Intrusion Recovery System}.
\newblock \bibinfo{journal}{\emph{SIGOPS Oper. Syst. Rev.}}
  (\bibinfo{year}{2005}).
\newblock


\bibitem[\protect\citeauthoryear{(GReAT)}{(GReAT)}{2015}]%
        {carbanak-report}
\bibfield{author}{\bibinfo{person}{Kaspersky Lab: Global Research \&
  Analysis~Team (GReAT)}.} \bibinfo{year}{2015}\natexlab{}.
\newblock \bibinfo{title}{{Carbanak APT: The Great Bank Robbery}}.
\newblock
  \bibinfo{howpublished}{\url{https://media.kasperskycontenthub.com/wp-content/uploads/sites/43/2018/03/08064518/Carbanak_APT_eng.pdf}}.
\newblock
\newblock
\shownote{Accessed: 2019-04-19.}


\bibitem[\protect\citeauthoryear{hasherezade}{hasherezade}{2018}]%
        {pe-sieve}
\bibfield{author}{\bibinfo{person}{hasherezade}.}
  \bibinfo{year}{2018}\natexlab{}.
\newblock \bibinfo{title}{{PE-Sieve}: Scans a given process. Recognizes and
  dumps a variety of potentially malicious implants (replaced/injected PEs,
  shellcodes, hooks, in-memory patches)}.
\newblock
  \bibinfo{howpublished}{\url{https://github.com/hasherezade/pe-sieve}}.
\newblock


\bibitem[\protect\citeauthoryear{Hassan, Guo, Li, Chen, Jee, Li, and
  Bates}{Hassan et~al\mbox{.}}{2019}]%
        {hassan2019nodoze}
\bibfield{author}{\bibinfo{person}{Wajih~Ul Hassan}, \bibinfo{person}{Shengjian
  Guo}, \bibinfo{person}{Ding Li}, \bibinfo{person}{Zhengzhang Chen},
  \bibinfo{person}{Kangkook Jee}, \bibinfo{person}{Zhichun Li}, {and}
  \bibinfo{person}{Adam Bates}.} \bibinfo{year}{2019}\natexlab{}.
\newblock \showarticletitle{NoDoze: Combatting Threat Alert Fatigue with
  Automated Provenance Triage.}. In \bibinfo{booktitle}{\emph{NDSS}}.
\newblock


\bibitem[\protect\citeauthoryear{Hossain, Milajerdi, Wang, Eshete, Gjomemo,
  Sekar, Stoller, and Venkatakrishnan}{Hossain et~al\mbox{.}}{2017}]%
        {hossain2017sleuth}
\bibfield{author}{\bibinfo{person}{Md~Nahid Hossain},
  \bibinfo{person}{Sadegh~M. Milajerdi}, \bibinfo{person}{Junao Wang},
  \bibinfo{person}{Birhanu Eshete}, \bibinfo{person}{Rigel Gjomemo},
  \bibinfo{person}{R. Sekar}, \bibinfo{person}{Scott Stoller}, {and}
  \bibinfo{person}{V.N. Venkatakrishnan}.} \bibinfo{year}{2017}\natexlab{}.
\newblock \showarticletitle{{SLEUTH}: Real-time Attack Scenario Reconstruction
  from {COTS} Audit Data}. In \bibinfo{booktitle}{\emph{26th {USENIX} Security
  Symposium ({USENIX} Security 17)}}. \bibinfo{publisher}{{USENIX}
  Association}, \bibinfo{address}{Vancouver, BC}, \bibinfo{pages}{487--504}.
\newblock
\showISBNx{978-1-931971-40-9}
\urldef\tempurl%
\url{https://www.usenix.org/conference/usenixsecurity17/technical-sessions/presentation/hossain}
\showURL{%
\tempurl}


\bibitem[\protect\citeauthoryear{Hossain, Wang, Sekar, and Stoller}{Hossain
  et~al\mbox{.}}{2018}]%
        {depPresRed18}
\bibfield{author}{\bibinfo{person}{Md~Nahid Hossain}, \bibinfo{person}{Junao
  Wang}, \bibinfo{person}{R. Sekar}, {and} \bibinfo{person}{Scott Stoller}.}
  \bibinfo{year}{2018}\natexlab{}.
\newblock \showarticletitle{Dependence Preserving Data Compaction for Scalable
  Forensic Analysis}. In \bibinfo{booktitle}{\emph{USENIX Security Symposium}}.
  \bibinfo{publisher}{{USENIX} Association}.
\newblock


\bibitem[\protect\citeauthoryear{Husari, Al-Shaer, Ahmed, Chu, and Niu}{Husari
  et~al\mbox{.}}{2017}]%
        {ttpdrill}
\bibfield{author}{\bibinfo{person}{Ghaith Husari}, \bibinfo{person}{Ehab
  Al-Shaer}, \bibinfo{person}{Mohiuddin Ahmed}, \bibinfo{person}{Bill Chu},
  {and} \bibinfo{person}{Xi Niu}.} \bibinfo{year}{2017}\natexlab{}.
\newblock \showarticletitle{TTPDrill: Automatic and Accurate Extraction of
  Threat Actions from Unstructured Text of CTI Sources}. In
  \bibinfo{booktitle}{\emph{Proceedings of the 33rd Annual Computer Security
  Applications Conference}}. \bibinfo{publisher}{ACM},
  \bibinfo{pages}{103--115}.
\newblock


\bibitem[\protect\citeauthoryear{Iklody}{Iklody}{2019}]%
        {misp-correlations}
\bibfield{author}{\bibinfo{person}{MISP:~Andras Iklody}.}
  \bibinfo{year}{2019}\natexlab{}.
\newblock \bibinfo{title}{{Default type of relationships in MISP objects.}}
\newblock
  \bibinfo{howpublished}{\url{https://github.com/MISP/misp-objects/blob/master/relationships/definition.json}}.
\newblock
\newblock
\shownote{Accessed: 2019-04-23.}


\bibitem[\protect\citeauthoryear{Ji, Lee, Downing, Wang, Fazzini, Kim, Orso,
  and Lee}{Ji et~al\mbox{.}}{2017}]%
        {ji2017rain}
\bibfield{author}{\bibinfo{person}{Yang Ji}, \bibinfo{person}{Sangho Lee},
  \bibinfo{person}{Evan Downing}, \bibinfo{person}{Weiren Wang},
  \bibinfo{person}{Mattia Fazzini}, \bibinfo{person}{Taesoo Kim},
  \bibinfo{person}{Alessandro Orso}, {and} \bibinfo{person}{Wenke Lee}.}
  \bibinfo{year}{2017}\natexlab{}.
\newblock \showarticletitle{Rain: Refinable Attack Investigation with On-demand
  Inter-Process Information Flow Tracking}. In
  \bibinfo{booktitle}{\emph{Proceedings of the 2017 ACM SIGSAC Conference on
  Computer and Communications Security}}. ACM, \bibinfo{pages}{377--390}.
\newblock


\bibitem[\protect\citeauthoryear{Ji, Lee, Fazzini, Allen, Downing, Kim, Orso,
  and Lee}{Ji et~al\mbox{.}}{2018}]%
        {ji2018enabling}
\bibfield{author}{\bibinfo{person}{Yang Ji}, \bibinfo{person}{Sangho Lee},
  \bibinfo{person}{Mattia Fazzini}, \bibinfo{person}{Joey Allen},
  \bibinfo{person}{Evan Downing}, \bibinfo{person}{Taesoo Kim},
  \bibinfo{person}{Alessandro Orso}, {and} \bibinfo{person}{Wenke Lee}.}
  \bibinfo{year}{2018}\natexlab{}.
\newblock \showarticletitle{Enabling refinable cross-host attack investigation
  with efficient data flow tagging and tracking}. In
  \bibinfo{booktitle}{\emph{27th $\{$USENIX$\}$ Security Symposium
  ($\{$USENIX$\}$ Security 18)}}. \bibinfo{pages}{1705--1722}.
\newblock


\bibitem[\protect\citeauthoryear{Keromytis}{Keromytis}{2018}]%
        {tc_github}
\bibfield{author}{\bibinfo{person}{Angelos~D. Keromytis}.}
  \bibinfo{year}{2018}\natexlab{}.
\newblock \bibinfo{title}{Transparent Computing Engagement 3 Data Release}.
\newblock
  \bibinfo{howpublished}{\url{https://github.com/darpa-i2o/Transparent-Computing}}.
\newblock


\bibitem[\protect\citeauthoryear{Khan, Wu, Aggarwal, and Yan}{Khan
  et~al\mbox{.}}{2013}]%
        {khan2013nema}
\bibfield{author}{\bibinfo{person}{Arijit Khan}, \bibinfo{person}{Yinghui Wu},
  \bibinfo{person}{Charu~C Aggarwal}, {and} \bibinfo{person}{Xifeng Yan}.}
  \bibinfo{year}{2013}\natexlab{}.
\newblock \showarticletitle{Nema: Fast graph search with label similarity}. In
  \bibinfo{booktitle}{\emph{Proceedings of the VLDB Endowment}},
  Vol.~\bibinfo{volume}{6}. VLDB Endowment, \bibinfo{pages}{181--192}.
\newblock


\bibitem[\protect\citeauthoryear{King and Chen}{King and Chen}{2003}]%
        {king2003backtracking}
\bibfield{author}{\bibinfo{person}{Samuel~T King} {and}
  \bibinfo{person}{Peter~M Chen}.} \bibinfo{year}{2003}\natexlab{}.
\newblock \showarticletitle{Backtracking intrusions}. In
  \bibinfo{booktitle}{\emph{SOSP}}. \bibinfo{publisher}{ACM}.
\newblock


\bibitem[\protect\citeauthoryear{King and Chen}{King and Chen}{2005}]%
        {king2005backtracking}
\bibfield{author}{\bibinfo{person}{Samuel~T. King} {and}
  \bibinfo{person}{Peter~M. Chen}.} \bibinfo{year}{2005}\natexlab{}.
\newblock \showarticletitle{Backtracking Intrusions}.
\newblock \bibinfo{journal}{\emph{ACM Transactions on Computer Systems}}
  (\bibinfo{year}{2005}).
\newblock


\bibitem[\protect\citeauthoryear{King, Mao, Lucchetti, and Chen}{King
  et~al\mbox{.}}{2005}]%
        {king2005enriching}
\bibfield{author}{\bibinfo{person}{Samuel~T King},
  \bibinfo{person}{Zhuoqing~Morley Mao}, \bibinfo{person}{Dominic~G Lucchetti},
  {and} \bibinfo{person}{Peter~M Chen}.} \bibinfo{year}{2005}\natexlab{}.
\newblock \showarticletitle{Enriching Intrusion Alerts Through Multi-Host
  Causality.}. In \bibinfo{booktitle}{\emph{NDSS}}.
\newblock


\bibitem[\protect\citeauthoryear{Kolbitsch, Comparetti, Kruegel, Kirda, Zhou,
  and Wang}{Kolbitsch et~al\mbox{.}}{2009}]%
        {kolbitsch2009effective}
\bibfield{author}{\bibinfo{person}{Clemens Kolbitsch},
  \bibinfo{person}{Paolo~Milani Comparetti}, \bibinfo{person}{Christopher
  Kruegel}, \bibinfo{person}{Engin Kirda}, \bibinfo{person}{Xiao-yong Zhou},
  {and} \bibinfo{person}{XiaoFeng Wang}.} \bibinfo{year}{2009}\natexlab{}.
\newblock \showarticletitle{Effective and Efficient Malware Detection at the
  End Host.}. In \bibinfo{booktitle}{\emph{USENIX security symposium}},
  Vol.~\bibinfo{volume}{4}. \bibinfo{pages}{351--366}.
\newblock


\bibitem[\protect\citeauthoryear{Kwon, Kim, Sumner, Kim, Saltaformaggio, Zhang,
  and Xu}{Kwon et~al\mbox{.}}{2016}]%
        {kwon2016ldx}
\bibfield{author}{\bibinfo{person}{Yonghwi Kwon}, \bibinfo{person}{Dohyeong
  Kim}, \bibinfo{person}{William~Nick Sumner}, \bibinfo{person}{Kyungtae Kim},
  \bibinfo{person}{Brendan Saltaformaggio}, \bibinfo{person}{Xiangyu Zhang},
  {and} \bibinfo{person}{Dongyan Xu}.} \bibinfo{year}{2016}\natexlab{}.
\newblock \showarticletitle{Ldx: Causality inference by lightweight dual
  execution}.
\newblock \bibinfo{journal}{\emph{ACM SIGOPS Operating Systems Review}}
  \bibinfo{volume}{50}, \bibinfo{number}{2} (\bibinfo{year}{2016}),
  \bibinfo{pages}{503--515}.
\newblock


\bibitem[\protect\citeauthoryear{Kwon, Wang, Wang, Lee, Lee, Ma, Zhang, Xu,
  Jha, Ciocarlie, et~al\mbox{.}}{Kwon et~al\mbox{.}}{2018}]%
        {kwon18mci}
\bibfield{author}{\bibinfo{person}{Yonghwi Kwon}, \bibinfo{person}{Fei Wang},
  \bibinfo{person}{Weihang Wang}, \bibinfo{person}{Kyu~Hyung Lee},
  \bibinfo{person}{Wen-Chuan Lee}, \bibinfo{person}{Shiqing Ma},
  \bibinfo{person}{Xiangyu Zhang}, \bibinfo{person}{Dongyan Xu},
  \bibinfo{person}{Somesh Jha}, \bibinfo{person}{Gabriela Ciocarlie},
  {et~al\mbox{.}}} \bibinfo{year}{2018}\natexlab{}.
\newblock \showarticletitle{MCI: Modeling-based Causality Inference in Audit
  Logging for Attack Investigation}. In \bibinfo{booktitle}{\emph{Proc. of the
  25th Network and Distributed System Security Symposium (NDSS’18)}}.
\newblock


\bibitem[\protect\citeauthoryear{Lee, Zhang, and Xu}{Lee
  et~al\mbox{.}}{2013a}]%
        {lee2013high}
\bibfield{author}{\bibinfo{person}{Kyu~Hyung Lee}, \bibinfo{person}{Xiangyu
  Zhang}, {and} \bibinfo{person}{Dongyan Xu}.}
  \bibinfo{year}{2013}\natexlab{a}.
\newblock \showarticletitle{High Accuracy Attack Provenance via Binary-based
  Execution Partition.}. In \bibinfo{booktitle}{\emph{NDSS}}.
\newblock


\bibitem[\protect\citeauthoryear{Lee, Zhang, and Xu}{Lee
  et~al\mbox{.}}{2013b}]%
        {lee2013loggc}
\bibfield{author}{\bibinfo{person}{Kyu~Hyung Lee}, \bibinfo{person}{Xiangyu
  Zhang}, {and} \bibinfo{person}{Dongyan Xu}.}
  \bibinfo{year}{2013}\natexlab{b}.
\newblock \showarticletitle{LogGC: garbage collecting audit log}. In
  \bibinfo{booktitle}{\emph{Proceedings of the 2013 ACM SIGSAC conference on
  Computer \& communications security}}. ACM, \bibinfo{pages}{1005--1016}.
\newblock


\bibitem[\protect\citeauthoryear{Liao, Yuan, Wang, Li, Xing, and Beyah}{Liao
  et~al\mbox{.}}{2016}]%
        {iace}
\bibfield{author}{\bibinfo{person}{Xiaojing Liao}, \bibinfo{person}{Kan Yuan},
  \bibinfo{person}{XiaoFeng Wang}, \bibinfo{person}{Zhou Li},
  \bibinfo{person}{Luyi Xing}, {and} \bibinfo{person}{Raheem Beyah}.}
  \bibinfo{year}{2016}\natexlab{}.
\newblock \showarticletitle{Acing the IOC game: Toward automatic discovery and
  analysis of open-source cyber threat intelligence}. In
  \bibinfo{booktitle}{\emph{Proceedings of the 2016 ACM SIGSAC Conference on
  Computer and Communications Security}}. ACM, \bibinfo{pages}{755--766}.
\newblock


\bibitem[\protect\citeauthoryear{Liu, Zhang, Li, Jee, Li, Wu, Rhee, and
  Mittal}{Liu et~al\mbox{.}}{2018}]%
        {liu2018towards}
\bibfield{author}{\bibinfo{person}{Yushan Liu}, \bibinfo{person}{Mu Zhang},
  \bibinfo{person}{Ding Li}, \bibinfo{person}{Kangkook Jee},
  \bibinfo{person}{Zhichun Li}, \bibinfo{person}{Zhenyu Wu},
  \bibinfo{person}{Junghwan Rhee}, {and} \bibinfo{person}{Prateek Mittal}.}
  \bibinfo{year}{2018}\natexlab{}.
\newblock \showarticletitle{Towards a Timely Causality Analysis for Enterprise
  Security}. In \bibinfo{booktitle}{\emph{Network and Distributed Systems
  Security Symposium}}.
\newblock


\bibitem[\protect\citeauthoryear{M.~Milajerdi, Eshete, Gjomemo, and
  Venkatakrishnan}{M.~Milajerdi et~al\mbox{.}}{2018}]%
        {sadegh2018propatrol}
\bibfield{author}{\bibinfo{person}{Sadegh M.~Milajerdi},
  \bibinfo{person}{Birhanu Eshete}, \bibinfo{person}{Rigel Gjomemo}, {and}
  \bibinfo{person}{V.N. Venkatakrishnan}.} \bibinfo{year}{2018}\natexlab{}.
\newblock \showarticletitle{ProPatrol: Attack Investigation via Extracted
  High-Level Tasks}. In \bibinfo{booktitle}{\emph{International Conference on
  Information Systems Security}}. Springer.
\newblock


\bibitem[\protect\citeauthoryear{Ma, Lee, Kim, Rhee, Zhang, and Xu}{Ma
  et~al\mbox{.}}{2015}]%
        {Ma2015Accurate}
\bibfield{author}{\bibinfo{person}{Shiqing Ma}, \bibinfo{person}{Kyu~Hyung
  Lee}, \bibinfo{person}{Chung~Hwan Kim}, \bibinfo{person}{Junghwan Rhee},
  \bibinfo{person}{Xiangyu Zhang}, {and} \bibinfo{person}{Dongyan Xu}.}
  \bibinfo{year}{2015}\natexlab{}.
\newblock \showarticletitle{Accurate, Low Cost and Instrumentation-Free
  Security Audit Logging for Windows}. In \bibinfo{booktitle}{\emph{Proceedings
  of the 31st Annual Computer Security Applications Conference}}
  \emph{(\bibinfo{series}{ACSAC 2015})}. \bibinfo{publisher}{ACM},
  \bibinfo{address}{New York, NY, USA}, \bibinfo{pages}{401--410}.
\newblock
\showISBNx{978-1-4503-3682-6}
\urldef\tempurl%
\url{https://doi.org/10.1145/2818000.2818039}
\showDOI{\tempurl}


\bibitem[\protect\citeauthoryear{Ma, Zhai, Wang, Lee, Zhang, and Xu}{Ma
  et~al\mbox{.}}{2017}]%
        {ma2017mpi}
\bibfield{author}{\bibinfo{person}{Shiqing Ma}, \bibinfo{person}{Juan Zhai},
  \bibinfo{person}{Fei Wang}, \bibinfo{person}{Kyu~Hyung Lee},
  \bibinfo{person}{Xiangyu Zhang}, {and} \bibinfo{person}{Dongyan Xu}.}
  \bibinfo{year}{2017}\natexlab{}.
\newblock \showarticletitle{MPI: Multiple Perspective Attack Investigation with
  Semantics Aware Execution Partitioning}. In \bibinfo{booktitle}{\emph{26th
  $\{$USENIX$\}$ Security Symposium ($\{$USENIX$\}$ Security 17)}}.
  \bibinfo{pages}{1111--1128}.
\newblock


\bibitem[\protect\citeauthoryear{Ma, Zhang, and Xu}{Ma et~al\mbox{.}}{2016}]%
        {ma2016protracer}
\bibfield{author}{\bibinfo{person}{Shiqing Ma}, \bibinfo{person}{Xiangyu
  Zhang}, {and} \bibinfo{person}{Dongyan Xu}.} \bibinfo{year}{2016}\natexlab{}.
\newblock \showarticletitle{{ProTracer}: {Towards} Practical Provenance Tracing
  by Alternating Between Logging and Tainting}. In
  \bibinfo{booktitle}{\emph{NDSS}}.
\newblock


\bibitem[\protect\citeauthoryear{Milajerdi, Gjomemo, Eshete, Sekar, and
  Venkatakrishnan}{Milajerdi et~al\mbox{.}}{2019}]%
        {sadegh2019holmes}
\bibfield{author}{\bibinfo{person}{Sadegh~M. Milajerdi}, \bibinfo{person}{Rigel
  Gjomemo}, \bibinfo{person}{Birhanu Eshete}, \bibinfo{person}{R. Sekar}, {and}
  \bibinfo{person}{VN. Venkatakrishnan}.} \bibinfo{year}{2019}\natexlab{}.
\newblock \showarticletitle{{HOLMES: Real-time APT Detection through
  Correlation of Suspicious Information Flows}}. In
  \bibinfo{booktitle}{\emph{Proceedings of the IEEE Symposium on Security and
  Privacy}}. \bibinfo{publisher}{IEEE}.
\newblock


\bibitem[\protect\citeauthoryear{MISP}{MISP}{2019}]%
        {misp}
\bibfield{author}{\bibinfo{person}{MISP}.} \bibinfo{year}{2019}\natexlab{}.
\newblock \bibinfo{title}{{MISP - Open Source Threat Intelligence Platform \&
  Open Standards For Threat Information Sharing}}.
\newblock \bibinfo{howpublished}{\url{https://www.misp-project.org/}}.
\newblock
\newblock
\shownote{Accessed: 2019-04-23.}


\bibitem[\protect\citeauthoryear{Mitre}{Mitre}{2018}]%
        {stix}
\bibfield{author}{\bibinfo{person}{Mitre}.} \bibinfo{year}{2018}\natexlab{}.
\newblock \bibinfo{title}{{Structured Threat Information eXpression (STIX)}}.
\newblock \bibinfo{howpublished}{\url{https://stixproject.github.io}}.
\newblock


\bibitem[\protect\citeauthoryear{Moran and Villeneuve}{Moran and
  Villeneuve}{2013}]%
        {deputydog-report}
\bibfield{author}{\bibinfo{person}{FireEye:~Ned Moran} {and}
  \bibinfo{person}{Nart Villeneuve}.} \bibinfo{year}{2013}\natexlab{}.
\newblock \bibinfo{title}{{Operation DeputyDog: Zero-Day (CVE-2013-3893) Attack
  Against Japanese Targets}}.
\newblock
  \bibinfo{howpublished}{\url{https://www.fireeye.com/blog/threat-research/2013/09/operation-deputydog-zero-day-cve-2013-3893-attack-against-japanese-targets.html}}.
\newblock
\newblock
\shownote{Accessed: 2019-04-19.}


\bibitem[\protect\citeauthoryear{Oprea, Li, Yen, Chin, and Alrwais}{Oprea
  et~al\mbox{.}}{2015}]%
        {oprea2015detection}
\bibfield{author}{\bibinfo{person}{Alina Oprea}, \bibinfo{person}{Zhou Li},
  \bibinfo{person}{Ting-Fang Yen}, \bibinfo{person}{Sang~H Chin}, {and}
  \bibinfo{person}{Sumayah Alrwais}.} \bibinfo{year}{2015}\natexlab{}.
\newblock \showarticletitle{Detection of early-stage enterprise infection by
  mining large-scale log data}. In \bibinfo{booktitle}{\emph{Dependable Systems
  and Networks (DSN), 2015 45th Annual IEEE/IFIP International Conference on}}.
  IEEE, \bibinfo{pages}{45--56}.
\newblock


\bibitem[\protect\citeauthoryear{Parampalli, Sekar, and Johnson}{Parampalli
  et~al\mbox{.}}{2008}]%
        {asiaccs08}
\bibfield{author}{\bibinfo{person}{Chetan Parampalli}, \bibinfo{person}{R
  Sekar}, {and} \bibinfo{person}{Rob Johnson}.}
  \bibinfo{year}{2008}\natexlab{}.
\newblock \showarticletitle{A practical mimicry attack against powerful
  system-call monitors}. In \bibinfo{booktitle}{\emph{Information, computer and
  communications security}}. ACM.
\newblock


\bibitem[\protect\citeauthoryear{Pasquier, Han, Moyer, Bates, Hermant, Eyers,
  Bacon, and Seltzer}{Pasquier et~al\mbox{.}}{2018}]%
        {pasquierruntime}
\bibfield{author}{\bibinfo{person}{Thomas Pasquier}, \bibinfo{person}{Xueyuan
  Han}, \bibinfo{person}{Thomas Moyer}, \bibinfo{person}{Adam Bates},
  \bibinfo{person}{Olivier Hermant}, \bibinfo{person}{David Eyers},
  \bibinfo{person}{Jean Bacon}, {and} \bibinfo{person}{Margo Seltzer}.}
  \bibinfo{year}{2018}\natexlab{}.
\newblock \showarticletitle{Runtime Analysis of Whole-System Provenance}. In
  \bibinfo{booktitle}{\emph{Proceedings of the 2018 ACM SIGSAC Conference on
  Computer and Communications Security}} \emph{(\bibinfo{series}{CCS '18})}.
  \bibinfo{publisher}{ACM}, \bibinfo{address}{New York, NY, USA},
  \bibinfo{pages}{1601--1616}.
\newblock
\showISBNx{978-1-4503-5693-0}
\urldef\tempurl%
\url{https://doi.org/10.1145/3243734.3243776}
\showDOI{\tempurl}


\bibitem[\protect\citeauthoryear{Pei, Gu, Saltaformaggio, Ma, Wang, Zhang, Si,
  Zhang, and Xu}{Pei et~al\mbox{.}}{2016}]%
        {pei2016hercule}
\bibfield{author}{\bibinfo{person}{Kexin Pei}, \bibinfo{person}{Zhongshu Gu},
  \bibinfo{person}{Brendan Saltaformaggio}, \bibinfo{person}{Shiqing Ma},
  \bibinfo{person}{Fei Wang}, \bibinfo{person}{Zhiwei Zhang},
  \bibinfo{person}{Luo Si}, \bibinfo{person}{Xiangyu Zhang}, {and}
  \bibinfo{person}{Dongyan Xu}.} \bibinfo{year}{2016}\natexlab{}.
\newblock \showarticletitle{Hercule: Attack story reconstruction via community
  discovery on correlated log graph}. In \bibinfo{booktitle}{\emph{Proceedings
  of the 32Nd Annual Conference on Computer Security Applications}}. ACM,
  \bibinfo{pages}{583--595}.
\newblock


\bibitem[\protect\citeauthoryear{Pienta, Tamersoy, Tong, and Chau}{Pienta
  et~al\mbox{.}}{2014}]%
        {pienta2014mage}
\bibfield{author}{\bibinfo{person}{Robert Pienta}, \bibinfo{person}{Acar
  Tamersoy}, \bibinfo{person}{Hanghang Tong}, {and} \bibinfo{person}{Duen~Horng
  Chau}.} \bibinfo{year}{2014}\natexlab{}.
\newblock \showarticletitle{Mage: Matching approximate patterns in
  richly-attributed graphs}. In \bibinfo{booktitle}{\emph{2014 IEEE
  International Conference on Big Data (Big Data)}}. IEEE,
  \bibinfo{pages}{585--590}.
\newblock


\bibitem[\protect\citeauthoryear{Pohly, McLaughlin, McDaniel, and Butler}{Pohly
  et~al\mbox{.}}{2012}]%
        {pohly2012hi}
\bibfield{author}{\bibinfo{person}{Devin~J Pohly}, \bibinfo{person}{Stephen
  McLaughlin}, \bibinfo{person}{Patrick McDaniel}, {and} \bibinfo{person}{Kevin
  Butler}.} \bibinfo{year}{2012}\natexlab{}.
\newblock \showarticletitle{Hi-Fi: collecting high-fidelity whole-system
  provenance}. In \bibinfo{booktitle}{\emph{ACSAC}}. ACM.
\newblock


\bibitem[\protect\citeauthoryear{Shu, Araujo, Schales, Stoecklin, Jang, Huang,
  and Rao}{Shu et~al\mbox{.}}{2018}]%
        {Shu:2018:TIC:3243734.3243829}
\bibfield{author}{\bibinfo{person}{Xiaokui Shu}, \bibinfo{person}{Frederico
  Araujo}, \bibinfo{person}{Douglas~L. Schales}, \bibinfo{person}{Marc~Ph.
  Stoecklin}, \bibinfo{person}{Jiyong Jang}, \bibinfo{person}{Heqing Huang},
  {and} \bibinfo{person}{Josyula~R. Rao}.} \bibinfo{year}{2018}\natexlab{}.
\newblock \showarticletitle{Threat Intelligence Computing}. In
  \bibinfo{booktitle}{\emph{Proceedings of the 2018 ACM SIGSAC Conference on
  Computer and Communications Security}} \emph{(\bibinfo{series}{CCS '18})}.
  \bibinfo{publisher}{ACM}, \bibinfo{address}{New York, NY, USA},
  \bibinfo{pages}{1883--1898}.
\newblock
\showISBNx{978-1-4503-5693-0}
\urldef\tempurl%
\url{https://doi.org/10.1145/3243734.3243829}
\showDOI{\tempurl}


\bibitem[\protect\citeauthoryear{Solutions}{Solutions}{2013}]%
        {njrat-report}
\bibfield{author}{\bibinfo{person}{General Dynamics Fidelis~Cybersecurity
  Solutions}.} \bibinfo{year}{2013}\natexlab{}.
\newblock \bibinfo{title}{{njRAT Uncovered}}.
\newblock
  \bibinfo{howpublished}{\url{https://app.box.com/s/vdg51zbfvap52w60zj0is3l1dmyya0n4}}.
\newblock
\newblock
\shownote{Accessed: 2019-04-19.}


\bibitem[\protect\citeauthoryear{Splunk}{Splunk}{2019}]%
        {splunk}
\bibfield{author}{\bibinfo{person}{Splunk}.} \bibinfo{year}{2019}\natexlab{}.
\newblock \bibinfo{title}{{SIEM}, {AIO}ps, {A}pplication {M}anagement, {L}og
  {M}anagement, {M}achine {L}earning, and {C}ompliance}.
\newblock \bibinfo{howpublished}{\url{https://www.splunk.com/}}.
\newblock


\bibitem[\protect\citeauthoryear{STIX}{STIX}{2019}]%
        {stix-visulaization}
\bibfield{author}{\bibinfo{person}{STIX}.} \bibinfo{year}{2019}\natexlab{}.
\newblock \bibinfo{title}{{STIX Visualization}}.
\newblock
  \bibinfo{howpublished}{\url{https://oasis-open.github.io/cti-documentation/stix/gettingstarted.html\#stix-visualization}}.
\newblock
\newblock
\shownote{Accessed: 2019-05-15.}


\bibitem[\protect\citeauthoryear{Sun, Dai, Liu, Singhal, and Yen}{Sun
  et~al\mbox{.}}{2018}]%
        {sun2018using}
\bibfield{author}{\bibinfo{person}{Xiaoyan Sun}, \bibinfo{person}{Jun Dai},
  \bibinfo{person}{Peng Liu}, \bibinfo{person}{Anoop Singhal}, {and}
  \bibinfo{person}{John Yen}.} \bibinfo{year}{2018}\natexlab{}.
\newblock \showarticletitle{Using Bayesian Networks for Probabilistic
  Identification of Zero-Day Attack Paths}.
\newblock \bibinfo{journal}{\emph{IEEE Transactions on Information Forensics
  and Security}} \bibinfo{volume}{13}, \bibinfo{number}{10}
  (\bibinfo{year}{2018}), \bibinfo{pages}{2506--2521}.
\newblock


\bibitem[\protect\citeauthoryear{Sun, Wang, Wang, Shao, and Li}{Sun
  et~al\mbox{.}}{2012}]%
        {sun2012efficient}
\bibfield{author}{\bibinfo{person}{Zhao Sun}, \bibinfo{person}{Hongzhi Wang},
  \bibinfo{person}{Haixun Wang}, \bibinfo{person}{Bin Shao}, {and}
  \bibinfo{person}{Jianzhong Li}.} \bibinfo{year}{2012}\natexlab{}.
\newblock \showarticletitle{Efficient subgraph matching on billion node
  graphs}.
\newblock \bibinfo{journal}{\emph{Proceedings of the VLDB Endowment}}
  \bibinfo{volume}{5}, \bibinfo{number}{9} (\bibinfo{year}{2012}),
  \bibinfo{pages}{788--799}.
\newblock


\bibitem[\protect\citeauthoryear{Symantec}{Symantec}{2019}]%
        {nsa-tools-china}
\bibfield{author}{\bibinfo{person}{Symantec}.} \bibinfo{year}{2019}\natexlab{}.
\newblock \bibinfo{title}{{Buckeye: Espionage Outfit Used Equation Group Tools
  Prior to Shadow Brokers Leak}}.
\newblock
  \bibinfo{howpublished}{\url{https://www.symantec.com/blogs/threat-intelligence/buckeye-windows-zero-day-exploit}}.
\newblock


\bibitem[\protect\citeauthoryear{Systems}{Systems}{2017}]%
        {loki}
\bibfield{author}{\bibinfo{person}{Nextron Systems}.}
  \bibinfo{year}{2017}\natexlab{}.
\newblock \bibinfo{title}{{LOKI}, free {IOC} scanner - {N}extron {S}ystems}.
\newblock \bibinfo{howpublished}{\url{https://www.nextron-systems.com/loki/}}.
\newblock


\bibitem[\protect\citeauthoryear{Team}{Team}{2016}]%
        {dustysky-report}
\bibfield{author}{\bibinfo{person}{ClearSky Cyber~Security Team}.}
  \bibinfo{year}{2016}\natexlab{}.
\newblock \bibinfo{title}{{Operation DustySky}}.
\newblock
  \bibinfo{howpublished}{\url{https://www.clearskysec.com/wp-content/uploads/2016/01/Operation\%20DustySky_TLP_WHITE.pdf}}.
\newblock
\newblock
\shownote{Accessed: 2019-04-19.}


\bibitem[\protect\citeauthoryear{team}{team}{2013a}]%
        {stix1}
\bibfield{author}{\bibinfo{person}{MITRE:~STIX team}.}
  \bibinfo{year}{2013}\natexlab{a}.
\newblock \bibinfo{title}{{APT1 Report Converstion to STIX}}.
\newblock
  \bibinfo{howpublished}{\url{https://stix.mitre.org/language/version1.0.1/samples/README.txt}}.
\newblock
\newblock
\shownote{Accessed: 2019-04-23.}


\bibitem[\protect\citeauthoryear{team}{team}{2013b}]%
        {stix2}
\bibfield{author}{\bibinfo{person}{MITRE:~STIX team}.}
  \bibinfo{year}{2013}\natexlab{b}.
\newblock \bibinfo{title}{{FireEye Poison Evy Report Converstion to STIX}}.
\newblock
  \bibinfo{howpublished}{\url{https://stix.mitre.org/language/version1.0.1/samples/README-fireeye.txt}}.
\newblock
\newblock
\shownote{Accessed: 2019-04-23.}


\bibitem[\protect\citeauthoryear{Times}{Times}{2019}]%
        {nsa-tools-china1}
\bibfield{author}{\bibinfo{person}{New~York Times}.}
  \bibinfo{year}{2019}\natexlab{}.
\newblock \bibinfo{title}{{How Chinese Spies Got the N.S.A.'s Hacking Tools,
  and Used Them for Attacks}}.
\newblock
  \bibinfo{howpublished}{\url{https://www.nytimes.com/2019/05/06/us/politics/china-hacking-cyber.html}}.
\newblock


\bibitem[\protect\citeauthoryear{Tong, Faloutsos, Gallagher, and
  Eliassi{-}Rad}{Tong et~al\mbox{.}}{2007}]%
        {gray}
\bibfield{author}{\bibinfo{person}{Hanghang Tong}, \bibinfo{person}{Christos
  Faloutsos}, \bibinfo{person}{Brian Gallagher}, {and} \bibinfo{person}{Tina
  Eliassi{-}Rad}.} \bibinfo{year}{2007}\natexlab{}.
\newblock \showarticletitle{Fast best-effort pattern matching in large
  attributed graphs}. In \bibinfo{booktitle}{\emph{13th {ACM} {SIGKDD}
  International Conference on Knowledge Discovery and Data Mining (KDD 2007)}}.
  \bibinfo{publisher}{{ACM}}, \bibinfo{pages}{737--746}.
\newblock


\bibitem[\protect\citeauthoryear{Wagner and Soto}{Wagner and Soto}{2002}]%
        {mimicry}
\bibfield{author}{\bibinfo{person}{David Wagner} {and} \bibinfo{person}{Paolo
  Soto}.} \bibinfo{year}{2002}\natexlab{}.
\newblock \showarticletitle{Mimicry attacks on host-based intrusion detection
  systems}. In \bibinfo{booktitle}{\emph{Proceedings of the 9th ACM Conference
  on Computer and Communications Security}}. ACM, \bibinfo{pages}{255--264}.
\newblock


\bibitem[\protect\citeauthoryear{Wang, Ding, Tung, Ying, and Jin}{Wang
  et~al\mbox{.}}{2012}]%
        {wang2012efficient}
\bibfield{author}{\bibinfo{person}{Xiaoli Wang}, \bibinfo{person}{Xiaofeng
  Ding}, \bibinfo{person}{Anthony~KH Tung}, \bibinfo{person}{Shanshan Ying},
  {and} \bibinfo{person}{Hai Jin}.} \bibinfo{year}{2012}\natexlab{}.
\newblock \showarticletitle{An efficient graph indexing method}. In
  \bibinfo{booktitle}{\emph{2012 IEEE 28th International Conference on Data
  Engineering}}. IEEE, \bibinfo{pages}{210--221}.
\newblock


\bibitem[\protect\citeauthoryear{Westcott and Bandla}{Westcott and
  Bandla}{2018}]%
        {aptnotes}
\bibfield{author}{\bibinfo{person}{David Westcott} {and} \bibinfo{person}{Kiran
  Bandla}.} \bibinfo{year}{2018}\natexlab{}.
\newblock \bibinfo{title}{{APT Notes}}.
\newblock \bibinfo{howpublished}{\url{https://github.com/aptnotes/data}}.
\newblock


\bibitem[\protect\citeauthoryear{Workbench}{Workbench}{2019}]%
        {jetstream}
\bibfield{author}{\bibinfo{person}{Workbench}.}
  \bibinfo{year}{2019}\natexlab{}.
\newblock \bibinfo{title}{{Jetstream2}}.
\newblock
  \bibinfo{howpublished}{\url{https://browserbench.org/JetStream/index.html}}.
\newblock
\newblock
\shownote{Accessed: 2019-08-27.}


\bibitem[\protect\citeauthoryear{Xu, Wu, Li, Jee, Rhee, Xiao, Xu, Wang, and
  Jiang}{Xu et~al\mbox{.}}{2016}]%
        {xu2016high}
\bibfield{author}{\bibinfo{person}{Zhang Xu}, \bibinfo{person}{Zhenyu Wu},
  \bibinfo{person}{Zhichun Li}, \bibinfo{person}{Kangkook Jee},
  \bibinfo{person}{Junghwan Rhee}, \bibinfo{person}{Xusheng Xiao},
  \bibinfo{person}{Fengyuan Xu}, \bibinfo{person}{Haining Wang}, {and}
  \bibinfo{person}{Guofei Jiang}.} \bibinfo{year}{2016}\natexlab{}.
\newblock \showarticletitle{High fidelity data reduction for big data security
  dependency analyses}. In \bibinfo{booktitle}{\emph{Proceedings of the 2016
  ACM SIGSAC Conference on Computer and Communications Security}}. ACM,
  \bibinfo{pages}{504--516}.
\newblock


\bibitem[\protect\citeauthoryear{Zhu and Dumitras}{Zhu and Dumitras}{2018}]%
        {zhu2018chainsmith}
\bibfield{author}{\bibinfo{person}{Ziyun Zhu} {and} \bibinfo{person}{Tudor
  Dumitras}.} \bibinfo{year}{2018}\natexlab{}.
\newblock \showarticletitle{Chainsmith: Automatically learning the semantics of
  malicious campaigns by mining threat intelligence reports}. In
  \bibinfo{booktitle}{\emph{2018 IEEE European Symposium on Security and
  Privacy (EuroS\&P)}}. IEEE, \bibinfo{pages}{458--472}.
\newblock


\bibitem[\protect\citeauthoryear{Zong, Raghavendra, Srivatsa, Yan, Singh, and
  Lee}{Zong et~al\mbox{.}}{2014}]%
        {zong2014cloud}
\bibfield{author}{\bibinfo{person}{Bo Zong}, \bibinfo{person}{Ramya
  Raghavendra}, \bibinfo{person}{Mudhakar Srivatsa}, \bibinfo{person}{Xifeng
  Yan}, \bibinfo{person}{Ambuj~K Singh}, {and} \bibinfo{person}{Kang-Won Lee}.}
  \bibinfo{year}{2014}\natexlab{}.
\newblock \showarticletitle{Cloud service placement via subgraph matching}. In
  \bibinfo{booktitle}{\emph{2014 IEEE 30th International Conference on Data
  Engineering}}. IEEE, \bibinfo{pages}{832--843}.
\newblock


\bibitem[\protect\citeauthoryear{Zong, Xiao, Li, Wu, Qian, Yan, Singh, and
  Jiang}{Zong et~al\mbox{.}}{2015}]%
        {zong2015behavior}
\bibfield{author}{\bibinfo{person}{Bo Zong}, \bibinfo{person}{Xusheng Xiao},
  \bibinfo{person}{Zhichun Li}, \bibinfo{person}{Zhenyu Wu},
  \bibinfo{person}{Zhiyun Qian}, \bibinfo{person}{Xifeng Yan},
  \bibinfo{person}{Ambuj~K Singh}, {and} \bibinfo{person}{Guofei Jiang}.}
  \bibinfo{year}{2015}\natexlab{}.
\newblock \showarticletitle{Behavior query discovery in system-generated
  temporal graphs}.
\newblock \bibinfo{journal}{\emph{Proceedings of the VLDB Endowment}}
  \bibinfo{volume}{9}, \bibinfo{number}{4} (\bibinfo{year}{2015}),
  \bibinfo{pages}{240--251}.
\newblock


\bibitem[\protect\citeauthoryear{Zou, Chen, and {\"O}zsu}{Zou
  et~al\mbox{.}}{2009}]%
        {zou2009distance}
\bibfield{author}{\bibinfo{person}{Lei Zou}, \bibinfo{person}{Lei Chen}, {and}
  \bibinfo{person}{M~Tamer {\"O}zsu}.} \bibinfo{year}{2009}\natexlab{}.
\newblock \showarticletitle{Distance-join: Pattern match query in a large graph
  database}.
\newblock \bibinfo{journal}{\emph{Proceedings of the VLDB Endowment}}
  \bibinfo{volume}{2}, \bibinfo{number}{1} (\bibinfo{year}{2009}),
  \bibinfo{pages}{886--897}.
\newblock


\end{thebibliography}

\appendix
\section{Appendix}\label{malware_description}
In this section, we provide a brief history of each malware and a summary of the statements  from their corresponding reports which we have used to construct the query graphs.

\noindent
\textbf{njRAT.}
njRAT is a publicly available Remote Access Trojan (RAT) that gives the attacker full control over the victim system.
Although the source code of njRAT is publicly available, attacks leveraging njRAT have mostly targeted organizations based in or focused on the Middle East region in the government, telecom, and energy sectors. 
When the malware is executed, 
it tries to read its configuration from a file with the extension of ``.exe.config'' (edge 1). %
njRAT malware stores the logged keystrokes in a ``.tmp'' file (edge 2), and also writes to  a 
``.pf'' file (edge 3).
To gain persistence, njRAT malware creates some copies of itself (edges 4\&8).
After execution (edges 5\&6), one of the copies writes to a ``.pf'' file (edge 7).
njRAT malware also start a netsh process 
located at 
(edge 9), 
which results in creation of 
another ``.pf'' file (edge 10).
Finally, the malware sets some registry values (edges 11-13) and beacons to a C2 server at 217.66.231.245 (edge 14).

\noindent
\textbf{DeputyDog.}
DeputyDog refers to a malware 
appearing to have targeted organizations in Japan, based on a report by FireEye.
The query graph that we extracted from the report of this malware is shown in \Cref{fig:deputydog}, and it is described in \cref{sec:approach_overview}.

\noindent
\textbf{Uroburos.}
Uroburos, ComRAT, Snake, Turla, and Agent.BTZ are all referring to a family of rootkit which is responsible for
the most significant breach of U.S. military computers.
The malware starts by dropping two  Microsoft Windows dynamic libraries
(edges 1\&2)
and calling rundll32.exe (edge 3) to install these libraries (edges 4\&5). 
Then, to be started during the boot process, the malware creates a registry key 
(edge 6).
The malware creates three log files 
(edges 7-9) and removes a set of file (edges 10-14).

\noindent
\textbf{Carbanak.}
Carbanak is a remote backdoor to provide remote access to infected machines.
The main motivation of the attackers appears to be financial gain, which has resulted in cumulative losses up to one billion dollars~\cite{carbanak-report}.
The compromise initially starts using a spear phishing email that appears to be legitimate banking communications (edge 1).
After the exploit, Carbanak copies itself into ``\%system32\%'' with the name ``svchost.exe'' (edges 2-4) and deletes the original file created by the exploit payload (edge 5).
To access autorun privileges, the malware creates a new service with a name in the format of ``<ServiceName>Sys'', where ServiceName is any existing service randomly chosen (edge 6).
Carbanak creates a file with a random name and a .bin extension
where it stores commands to be executed (edge 7).
Then, the malware gets the proxy configuration from a registry entry
(edge 8) and the Mozilla Firefox configuration file
(edge 9).
Finally, Carbanak communicates with its C2 server (edge 10).

\noindent
\textbf{DustySky.}
DustySky is a multi-stage malware whose main objective is intelligence gathering for political purposes.
The malware sample is disguised as a Microsoft Word file, and once it is executed (edge 1), a lure Microsoft word document in the Arabic language is opened (edges 2\&3) while the malware performs intelligence gathering in the background.
For VM evasion, the dropper checks the existence of some DLL files, specifically vboxmrxnp.dll and vmbusres.dll which indicate existence of VirtualBox (edges 4\&5) and vmGuestlib.dll which indicates existence of VMware (edge 6). 
DustySky Core is dropped to \%TEMP\% (edges 7\&8\&9), and keystroke logs are saved to \%TEMP\%\textbackslash temps (edge 10).

\noindent
\textbf{OceanLotus.}
OceanLotus, also known as APT32,
is believed to be a Vietnam-based APT group
targeting Southeast Asian countries.
After execution of this malware (edge 1), a decoy document and an eraser application are dropped (edges 2\&3), and the decoy document is lunched in Microsoft Word (edges 4\&5).
Then, the executable decrypts its resources and drops a copy of legitimate Symantec Network Access Control application (edge 6), an encrypted backdoor (edge 7), and a malicious DLL file (edge 8).
The Symantec application, which is signed and legitimate, loads all the libraries in the same folder by default. In this case, after execution (edges 9\&10), this application loads the malicious DLL file  which has been dropped in the same directory (edge 11).
It then reads the backdoor file (edge 12) which results in accessing a registry
(edge 13), loading the HTTPProv.dll library (edge 14), and creating a registry key
(edge 15).
Finally, the malware connects to its mothership (edges 16\&17).

\noindent
\textbf{HawkEye.}
HawkEye is a malware-as-a-service credential stealing malware  and is a popular tool among APT campaigns.
The new variant of this malware uses process hollowing to inject its code into the legitimate signed .NET framework executables and ships with many sophisticated functions to evade detection.
This new variant is usually delivered as a compressed file, and after decompression (edges 1\&2) and execution(edge 3), it spawns a child process (edge 4), called RegAsm, which is an assembly registration tool from the Microsoft .Net framework. 
HawkEye extracts a PE file into its memory and then injects it into the RegAsm process.
After sleeping for 10 seconds, the RegAsm process spawns two child processes named vbc both from the .Net framework as well (edges 5\&6).
One of these processes collects credentials of browsers, while the other one focuses on email and Instant Messaging (IM) appllications.
We have added one node, typed as a file or registry, corresponding to the name of each browser (edges 7-18) or email/IM (edges 19-26) application mentioned in the report.
Note that these applications might store some confidential information of interest to attackers into both files or registries, and that is why we did not limit our search to only files or registries.
The collected credentials are regularly saved into $*$.tmp files in the \%temp\% directory (edges 27\&28), while after a while, the RegAsm process reads the entire data of these tmp files into its memory (edges 29\&30) and deletes them immediately (edges 31\&32). 
Finally, RegAsm looks up the machines public IP from ``http[s]:\textbackslash\textbackslash whatismyipaddress.com\textbackslash'' web service (edges 33\&34) and then exfiltrates the collected information to the attacker's email address (edge 35).

It is important to note that there are some nodes with exactly same label and type in the query graph of HawkEye, such as F\&G or J\&K. However, these nodes get aligned to different nodes based on their dependencies with other entities. For example, node F interacts with browser applications while node G interacts with the email/IM applications.
In addition, the alignment of browser or mail application nodes is independent of their installation on the system. Many of these applications are not installed on the test machine, however when the malware attempts to check whether these applications are installed on the system, it initiates an OPEN event which gets detected by \projname.

\end{document}